\documentclass[12pt, draftclsnofoot, onecolumn]{IEEEtran}
\usepackage{amsfonts}
\usepackage{latexsym,bm}
\usepackage{amsmath,amssymb,amsfonts}
\usepackage{indentfirst}
\usepackage{cite}
\usepackage{graphicx,color,psfrag,overpic,subfigure}
\usepackage{stfloats}
\usepackage{cases}
\usepackage{array}
\usepackage{verbatim}
\usepackage{setspace}
\usepackage{fancyhdr}
\usepackage{multirow}

\newtheorem{prop}{Proposition}
\newtheorem{lemma}{Lemma}
\newtheorem{remark}{Remark}

\newtheorem{corollary}{Corollary}

\ifCLASSINFOpdf
\else
\fi

\begin{document}

%
% paper title
% can use linebreaks \\ within to get better formatting as desired
\title{A Data-Aided Channel Estimation Scheme for Decoupled Systems in Heterogeneous Networks}

% author names and affiliations
% use a multiple column layout for up to three different
% affiliations
\author{Wen~Liu,~\IEEEmembership{Student Member,~IEEE,}
        Kai-Kit~Wong,~\IEEEmembership{Fellow,~IEEE,} \\
        Shi~Jin,~\IEEEmembership{Senior Member,~IEEE,} and~Xiaohu You,~\IEEEmembership{Fellow,~IEEE}
\thanks{W. Liu, S. Jin and X. You are with the National Mobile Communications Research
Laboratory, Southeast University, Nanjing 210096, China (e-mail: newen@seu.edu.cn; jinshi@seu.edu.cn; xhyu@seu.edu.cn).

K.-K. Wong is with the Department of Electronic and Electrical Engineering, University College London, London WC1E 7JE, U.K. (e-mail: kai-kit.wong@ucl.ac.uk).}}

\maketitle

\begin{abstract}
Uplink/downlink (UL/DL) decoupling promises more flexible cell association and higher throughput in heterogeneous networks (HetNets), however, it hampers the acquisition of DL channel state information (CSI) in time-division-duplex (TDD) systems due to different base stations (BSs) connected in UL/DL. In this paper, we propose a novel data-aided (DA) channel estimation scheme to address this problem by utilizing decoded UL data to exploit CSI from received UL data signal in decoupled HetNets where a massive multiple-input multiple-output BS and dense small cell BSs are deployed. We analytically estimate BER performance of UL decoded data, which are used to derive an approximated normalized mean square error (NMSE) expression of the DA minimum mean square error (MMSE) estimator. Compared with the conventional least square (LS) and MMSE, it is shown that NMSE performances of all estimators are determined by their signal-to-noise ratio (SNR)-like terms and there is an increment consisting of UL data power, UL data length and BER values in the SNR-like term of DA method, which suggests DA method outperforms the conventional ones in any scenarios. Higher UL data power, longer UL data length and better BER performance lead to more accurate estimated channels with DA method. Numerical results verify that the analytical BER and NMSE results are close to the simulated ones and a remarkable gain in both NMSE and DL rate can be achieved by DA method in multiple scenarios with different modulations.
\end{abstract}

\begin{IEEEkeywords}
 Channel estimation, data-aided, decoupled system, heterogeneous network, massive MIMO.
\end{IEEEkeywords}

\section{Introduction}
Heterogeneous networks (HetNets) and massive multiple-input multiple-output (MIMO) are regarded as two key technologies for 5G, promising higher coverage and spectral efficiency \cite{marzetta2010noncooperative,boccardi2014five,andrews2014will,wang2014cellular}. To meet the demands of 1000$\times$ increase of throughput and 1000$\times$ less energy consumption at the same time\cite{qual2012data1000,huawei2013data1000}, massive MIMO base station (BS) is recently introduced to HetNets, where small cells such as pico- or femto-cells are densely deployed. Such model is referred to as massive MIMO HetNets \cite{andrews2013seven,adhikary2015massive,hoydis2013making,bjornson2013massive}. It is anticipated that the massive MIMO HetNet architecture will greatly improve throughput and regional coverage, and is also desirable for interference management and energy efficiency\cite{liu2013massive,lu2014overview}.

Different from homogeneous networks, cell association in HetNets is a tricky business to deal with. From the perspective of user equipment (UE), the BS with maximum downlink (DL) power is probably not the one with maximum uplink (UL) power. In \cite{andrews2013seven}, the notion of decoupling UL and DL was first introduced and UEs can connect to the BSs with highest signal-to-noise ratio (SNR) in both directions separately. With such flexible cell association policy, throughput and coverage probability are found to be much improved especially for cell edge UEs by theoretical analysis and simulation results in \cite{smiljkovikj2015analysis}. Similar positive results of load balancing and energy efficiency were presented in \cite{singh2015joint,elshaer2015load,smiljkovikj2015efficiency}. Hence, prior literature has adopted decoupling access as a strong candidate in the next generation network, though there are still major problems to be solved before facilitating this structure, for instance, channel estimation at the DL BS, DL precoding, signal synchronization, offloading techniques and so on \cite{boccardi2016decouple}.

Channel estimation is an important issue in wireless communication and also considered as a major bottleneck for massive MIMO systems \cite{marzetta2010noncooperative,ngo2013energy,yin2013coordinated}. A standard approach in time-division-duplex (TDD) massive MIMO systems is to exploit channel reciprocity such that channels are estimated by pilot-based training in UL and then treated as the real channels for DL beamforming. Unfortunately, channel reciprocity cannot be used directly in decoupled HetNets since UEs may associate with different BSs in UL/DL. Although DL BS can estimate DL channels of decoupled UEs from the received UL training signal, but it should be noted that unlike coupled UEs, decoupled UEs are those who are neither close to its UL BS nor DL BS, and can hardly have sound channel estimation performance.

This motivates us to propose a data-aided channel estimation scheme to enable decoupling access in TDD systems, especially for massive MIMO where a large number of channel elements have to be estimated. Using decoded data to aid channel estimation has been studied in \cite{ma2014dataConf,baltersee2001achievable,ma2014dataJouranl,khalighi2006data,coldrey2008training,jindal2009value,schoeneich2006iterative,zhao2008iterative} before. In \cite{coldrey2008training,jindal2009value}, the mutual information and a capacity lower bound for data-aided single-user MIMO systems were investigated and it was shown that data-aided methods permit the use of a very small number of pilots to achieve high spectral efficiency. An iterative joint channel estimation and data detection process was also investigated in \cite{ma2014dataConf,ma2014dataJouranl,schoeneich2006iterative,zhao2008iterative}, fidning that data-aided methods can effectively suppress the contamination effect in large-scale antenna systems \cite{ma2014dataConf,ma2014dataJouranl}. Nevertheless, prior works mainly focused on homogeneous networks and it is of significant importance to bring this idea into UL/DL decoupled HetNets for more efficient and reliable channel acquisition. To the best of authors' knowledge, it is the first time that channel estimation problem is discussed in cellular HetNets with decoupling access.

In this paper, we consider a single-cell HetNet with decoupling access where a macro base station (MBS) is in the center and small cell BSs (SBSs) are randomly but densely populated within the cell. A novel three-stage data-aided scheme is proposed to solve the DL channel acquisition problem of decoupled UEs at MBS\footnote{In two-tier HetNets, decoupled UEs always connect to an SBS in UL and an MBS in DL due to the user association policy.}. To implement the data-aided scheme, once data detection is finished at SBSs, the decoded data sequences and estimated bit error ratio (BER) values are supposed to be sent to the MBS via wired backhaul which is assumed to be error-free and latency-free. Then, the MBS utilizes known training sequences along with these decoded sequences and BER values to recover channels of decoupled UEs from received training and UL data signals.
%We propose a novel three-stage data-aided scheme to solve the channel estimation problem for decoupled UEs. The estimation process takes place in a few stages. In the first stage, all UEs send prescribed orthogonal training sequences to their serving SBSs simultaneously. Then the UL training signal is received and proceeded to the channel estimator at each SBS while the MBS\footnote{In two-tier HetNets, decoupled UEs always connect to an SBS in UL and an MBS in DL due to the user association policy.} needs to store the received UL training signal for subsequent estimation. In the second stage, the SBSs receive the UL data signals from all the UEs and try to recover the data of the UEs they serve. The UL data signal is also recorded and stored at the MBS. At the same time, we calculate the average bit error ratio (BER) of each data stream by averaging over fast fading. Then the decoded data along with the BER value are sent to the MBS via an error-free latency-free backhaul link. In the last stage, the decoupled UEs' channels are estimated at the MBS from the combined training signals and data signals with the knowledge of training sequences and the decoded data sequences.
The core idea is to employ the decoded data as extended training sequences to exploit the most channel information from received UL signals although the decoded data is not orthogonal and subject to unknown errors. Different from the multi-cell model with one UE in each cell discussed in \cite{ma2014dataConf,ma2014dataJouranl}, we consider a single-cell model\footnote{Here, single-cell represents single macro-cell, yet there are multiple overlapping small cells actually. This model can be generalized to multi-cell scenarios straightforwardly.} with dense small cells and multiple UEs where interference is much more severe. Furthermore, in \cite{ma2014dataConf,ma2014dataJouranl}, Gaussian data was assumed to be received at the least-square (LS) data estimator and the data estimation error was modelled as a Gaussian variable while in our scheme, coded data is detected by a minimum mean-square-error (MMSE) data estimator and the BER is estimated by analytical derivation. However, similar conclusions are drawn that estimation performance can be greatly improved and both co-channel interference and detection error could seriously compromise performance.

More specifically, our main contribution is to develop a novel data-aided channel estimation scheme for decoupled UEs in cellular decoupled HetNets by using decoded UL data with consideration of BER. We calculate the BER of BPSK-modulated UL data after the MMSE decoder with imperfect channel state information (CSI) and use this estimated BER to derive a closed-form approximated normalized mean-square-error (NMSE) expression for the proposed data-aided scheme. We further compare NMSE performance between conventional LS, MMSE and data-aided MMSE and find NMSE expressions of all estimators have the same structure and are commonly determined by an similar SNR-like term, however, compared with conventional estimators, there is an additional part consisting of UL data power, UL data length and BER values in the SNR-like term of data-aided scheme, which explicitly explains how NMSE performance benefits from the data-aided method. Average DL rate performance is also analyzed numerically with NLoS and NLoS/LoS models to verify the improvement in final DL performance and 4QAM/16QAM are utilized to suggest that the proposed scheme can be applied in multiple scenarios with different modulations.

%Moreover, the NMSE expression of data-aided scheme shows that there is a performance floor when increasing UL data power in the scenario with BER while NMSE performance is determined by the total energy of training and UL data slots in the error-free scenario.

The reminder of this paper is organized as follows. Section \uppercase\expandafter{\romannumeral2} introduces the system model of decoupled HetNets. A novel data-aided scheme is proposed and elaborated in Section \uppercase\expandafter{\romannumeral 3}. The proposed scheme is compared with conventional channel estimation methods under the evaluation criteria of NMSE in Section \uppercase\expandafter{\romannumeral 4}. The theoretical results and some insights into this scheme are also discussed. Numerical results are presented in Section \uppercase\expandafter{\romannumeral 5} and Section \uppercase\expandafter{\romannumeral 6} concludes the paper.

\section{System Model}
\begin{figure}[!htbp]
\centering
	\includegraphics[width=10cm]{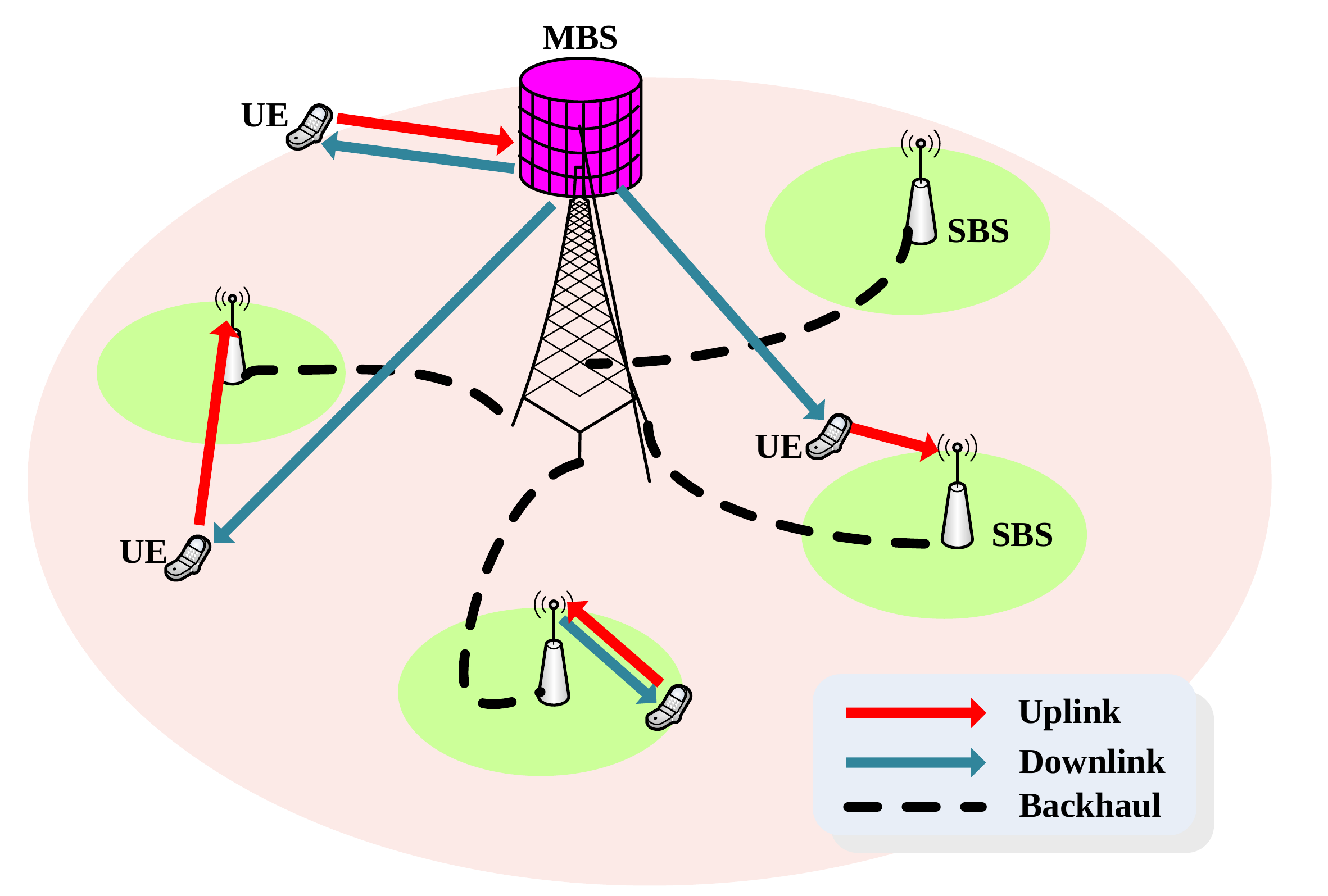}
    \caption{Diagram for cellular HetNets with decoupled access.}
    \label{Decoupled_Fiber}
\end{figure}

We consider a single-cell scenario deployed with an MBS in the center and $S$ SBSs along with $K$ UEs scattering in the range of the cell randomly as shown in Fig.~\ref{Decoupled_Fiber}. For ease of geometric analysis, we assume the single cell as a circular area with radius of $R_{\rm{M}}$. Meanwhile, the MBS equipped with $M$ antennas and all SBSs with $N$ antennas provide services to all single-antenna UEs in the cell coverage by fully utilizing the whole frequency band without any partitions, meaning that this model is interference limited. All communication links in the system are operating in TDD mode. Note that in this paper, we assume that there exist capacity-abundant fibers for backhaul between all SBSs and the MBS, which allows decoded data to be transmitted to the MBS without any latency nor error.

We assume that the channel coherence time is $T$, within which channel estimation and UL/DL transmission are processed. The channel from UEs to the MBS is denoted by ${\bf{H}} \in \mathbb{C}^{M \times K}$ while the channel from UEs to the $s$th SBS is represented by ${\bf{G}}_s \in \mathbb{C}^{N \times K}$. Also, ${\bf{h}}_k$, ${\bf{g}}_{sk}$ are the $k$th columns of ${\bf{H}}$ and ${\bf{G}}_s$, respectively, which are of the forms:
\begin{equation}\label{}
\begin{array}{l}
{{\bf{h}}_k} = \sqrt {\beta _k^{\rm{M}}} {\bf{h}}_k^{\rm{W}},\\
{{\bf{g}}_{sk}} = \sqrt {\beta _{sk}^{\rm{S}}} {\bf{g}}_{sk}^{\rm{W}},
\end{array}
\end{equation}
where each element of ${\bf{h}}_k^{\rm{W}} \in \mathbb{C}^{M \times 1}$ and ${\bf{g}}_{sk}^{\rm{W}} \in \mathbb{C}^{N \times 1} $ follows $\mathcal{CN}(0,1)$, ${\beta _k^{\rm{M}}}$ and ${\beta _{sk}^{\rm{S}}}$ represent the large scale fading from the $k$th UE to the MBS and the $s$th SBS, respectively, by neglecting shadowing effect and differences among antennas. The large scale fading between the $k$th UE to the MBS is modelled as $\beta _k^{\rm{M}} = {\left( {d_k^{\rm{M}}} \right)^{ - \alpha }}$, where ${d_k^{\rm{M}}}$ is the distance with attenuation exponent $\alpha$, and ${\beta _{sk}^{\rm{S}}}$ is defined similarly.

Unlike maximum average downlink receive power (MARP) \cite{elshaer2014downlink,smiljkovikj2015efficiency,smiljkovikj2015analysis} and biased cell association\cite{singh2015joint,jo2012heterogeneous} policies, due to the spectacular disparity between MBS and SBS, a new association strategy taking beamforming gain and transmit power into account is adopted in this paper. In particular, UEs perform a modified MARP policy in UL and DL respectively to achieve optimal associations in both links.

The received signal power in DL at the $k$th UE from a BS is given by
\begin{equation}\label{}
Q_{vk}^{{\rm{DL}}} =
\begin{cases}
{{{P_{\rm{M}}}}{{\left\| {{{\bf{h}}_k}} \right\|}^2},} & {v = 0},\\
{{{P_{\rm{S}}}}{{\left\| {{{\bf{g}}_{vk}}} \right\|}^2}},& {v = 1,2, \dots ,S},
\end{cases}
\end{equation}
where ${P_{\rm{M}}}$ and ${P_{\rm{S}}}$ are the total transmit power of MBS and SBS. In addition, the subscript $v$ represents the index of BS,  $0$ is for the MBS and $v$ is for the $v$th SBS. Similarly, the received signal power in UL at a BS from UE $k$ can be expressed as
\begin{equation}\label{}
Q_{vk}^{{\rm{UL}}} =
\begin{cases}
{{P_{\rm{D}}}{{\left\| {{{\bf{h}}_k}} \right\|}^2}}, & {v = 0},\\
{{P_{\rm{D}}}{{\left\| {{{\bf{g}}_{vk}}} \right\|}^2}},& {v = 1,2, \dots ,S},
\end{cases}
\end{equation}
where ${P_{\rm{D}}}$ stands for the transmit power of UEs.

Like the conventional MARP strategy, the effect of fast-fading should be averaged to let long-term parameters determine the association, which is more practical and operational in practice. Therefore, we exhibit the association results of the $k$th UE as an example using this modified MARP strategy:
\begin{equation}\label{}
\begin{array}{l}
{{\mathcal D}_k} = \mathop {\arg \max }\limits_{v = 0,1, \dots ,S} {\rm{ }}\left\{ {\mathop \mathbb{E}\limits_{_{{\bf{h}}_k^{\rm{W}},{\bf{g}}_{sk}^{\rm{W}}}} \left[ {Q_{vk}^{{\rm{DL}}}} \right]} \right\} = 
\mathop {\arg \max }\limits_{v = 0,1, \dots ,S} \left\{ {M{P_{\rm{M}}}{{\left( {d_k^{\rm{M}}} \right)}^{ - \alpha }},N{P_{\rm{S}}}{{\left( {d_{1k}^{\rm{S}}} \right)}^{ - \alpha }}, \dots ,N{P_{\rm{S}}}{{\left( {d_{Sk}^{\rm{S}}} \right)}^{ - \alpha }}} \right\},\\
{{\mathcal U}_k} = \mathop {\arg \max }\limits_{v = 0,1, \dots ,S} \left\{ {\mathop \mathbb{E} \limits_{_{{\bf{h}}_k^{\rm{W}},{\bf{g}}_{sk}^{\rm{W}}}} \left[ {Q_{vk}^{{\rm{UL}}}} \right]} \right\} = 
\mathop {\arg \max }\limits_{v = 0,1, \dots ,S} \left\{ {M{P_{\rm{D}}}{{\left( {d_k^{\rm{M}}} \right)}^{ - \alpha }},N{P_{\rm{D}}}{{\left( {d_{1k}^{\rm{S}}} \right)}^{ - \alpha }}, \dots ,N{P_{\rm{D}}}{{\left( {d_{Sk}^{\rm{S}}} \right)}^{ - \alpha }}} \right\}
\end{array}
\end{equation}
where ${\cal D}=\left\{{{\cal D}_1},\dots ,{{\cal D}_K}\right\}$ and ${\cal U}=\left\{{{\cal U}_1},\dots ,{{\cal U}_K}\right\}$ are DL and UL association sets. Based on this, we can focus on the UEs with indices belonging to the set $\left\{k| {{\cal D}_k} \ne {{\cal U}_k}, k=1,\dots,K \right\}$, who are regarded as decoupled UEs.

After performing the modified MARP strategy, UEs are able to connect to the optimal BSs in both links and ready to communicate with them. However, in decoupled HetNets, CSI acquired by UL training at the DL BS is not good enough to perform accurate DL precoding since the DL BS may not be the optimal UL BS. For the rest of this paper, we are devoted to improve the accuracy of DL channel for decoupled UEs by using decoded UL data to aid the estimation.

\section{Three-Stage Data-Aided Channel Estimation}
\begin{figure}
\centering
\includegraphics[width=10cm]{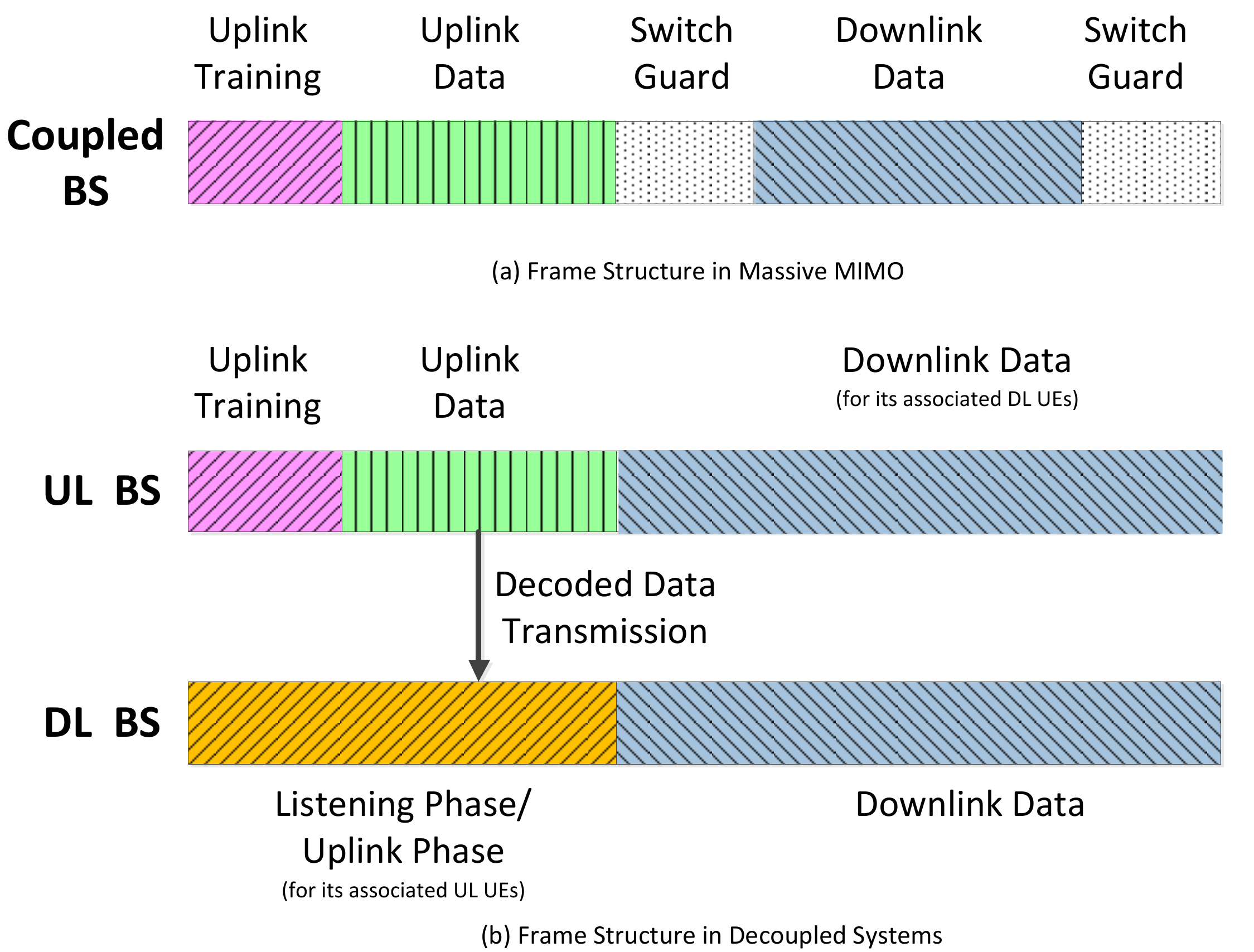}
    \caption{The frame structure for the data-aided channel estimation scheme.}
    \label{FrameStructure}
\end{figure}

In this section, we elucidate our proposed data-aided channel estimation scheme for cellular HetNets with decoupled access. The whole process can be described in three stages. To illustrate the sequence of operations in each stage, the frame structure of this data-aided scheme is shown in Fig.~\ref{FrameStructure}. Different from conventional frame structure of a massive MIMO system, UL BSs in decoupled systems need to transmit decoded data to DL BS after completing common UL training and UL data transmission, and the DL BS is required to listen and record the UL signal (including training and data phases) to perform joint channel estimation along with known training sequences and decoded data. Note that switch guard is ignored in decoupled systems for ease of understanding. The detailed implementation and signal model of the three-stage data-aided scheme are described step by step as follows.

\subsection{Stage 1: Uplink Training}
In the first stage, all UEs transmit their prescribed pilot sequences at power of ${{P_{\rm{T}}}}$. The pilot sequences are orthogonal with $\tau_{\rm{T}}$ symbols and ${\bf{S}} = {\left[ {{\bf{s}}_1^{\rm{T}}, \cdots {\bf{s}}_k^{\rm{T}}, \cdots ,{\bf{s}}_K^{\rm{T}}} \right]^{\rm{T}}}$ denotes the pilot matrix where ${\bf{s}}_k  \in {\mathbb{C}^{1 \times \tau_{\rm{T}} }}$ is the pilot for the $k$th UE. Obviously, we have ${\bf{S}}{{\bf{S}}^{\rm{H}}} = \tau_{\rm{T}}{{P_{\rm{T}}}}{{\bf{I}}_K}$ and $\tau_{\rm{T}} \ge K$. The main task here is for all UL BSs to recover channels from the UL training signal, which is a common procedure in training based systems. To this end, we consider two conventional channel estimators, LS and MMSE.

The received signal at the $v$th SBS can be expressed as
\begin{equation}\label{}
   {{\bf{Y}}_v^{{\rm{P}}}} =  {{\bf{G}}_v}{\bf{S}} + {{\bf{N}}_v} = \sum\limits_{k = 1}^K {{{\bf{g}}_{vk}}{{\bf{s}}_k}}  + {{\bf{N}}_v},
\end{equation}
where ${{\bf{N}}_v}\in\mathbb{C}^{N \times \tau_{\rm{T}}}$ denotes the additive white Gaussian noises (AWGNs) received and each element is independent identical distributed (i.i.d.), subject to $\mathcal{CN}(0,N_0)$.

Adopting the LS channel estimator, the estimated channel of the $k$th decoupled UE connecting to the $v$th SBS in UL is given as
\begin{equation}\label{EstChSC}
    {\bf{\hat g}}_{vk}^{{\rm{LS}}} = {{\bf{Y}}_v^{\rm{P}}}{\bf{s}}_k^{\rm{H}}{\left( {{{\bf{s}}_k}{\bf{s}}_k^{\rm{H}}} \right)^{ - 1}} = {{\bf{g}}_{vk}} + \frac{{{\bf{N}}_v}{\bf{s}}_k^{\rm{H}}}{\tau_{\rm{T}}{{P_{\rm{T}}}}}.
\end{equation}\vspace{-20pt}

MMSE is another widely used channel estimator which provides better performance at the cost of higher complexity and prior statistics information in terms of correlation matrices of channels and noises. MMSE estimator is served as a benchmark and the estimated channel is
\begin{equation}\label{EstChMMSE}
  {\bf{\hat g}}_{vk}^{{\rm{MMSE}}} = {{\bf{Y}}_v^{\rm{P}}}{\bf{C}}_{vk}^{{\rm{UP,opt}}} = {{\bf{G}}_v}{\bf{SC}}_{vk}^{{\rm{UP,opt}}} + {{\bf{N}}_v}{\bf{C}}_{vk}^{{\rm{UP,opt}}},
\end{equation}
\vspace{-5pt}where ${\bf{C}}_{vk}^{{\rm{UP,opt}}}$ is the ${{\tau _{\rm{T}}}} \times 1 $ linear estimation matrix for ${{{\bf{g}}_{vk}}}$, defined as
\begin{equation}\label{MMSECombination}
\begin{split}
  {\bf{C}}_{vk}^{{\rm{UP,opt}}}
\triangleq \arg \mathop {\min }\limits_{{{\bf{C}}_{vk}}} \mathbb{E} \left[ {{{\left\| {{{\bf{g}}_{vk}} - {\bf{\hat g}}_{vk}^{{\rm{MMSE}}}} \right\|}^2}} \right] = {\left[ {\sum\limits_{i = 1}^K {{\bf{s}}_i^{\rm{H}}{{{\bf R}}_{{{\bf{g}}_{vi}}}}{{\bf{s}}_i} + {{\bf{R}}_{{{\bf{N}}_v}}}} } \right]^{ - 1}}{\bf{s}}_k^{\rm{H}}{{{\bf R}}_{{{\bf{g}}_{vk}}}} \\
%  &= {\left[ {\sum\limits_{i = 1}^K {\beta _{vi}^{\rm{S}}{\bf{s}}_i^{\rm{H}}{{\bf{s}}_i} + {N_0}{{\bf{I}}_{{\tau _{\rm{T}}}}}} } \right]^{ - 1}}\beta _{vk}^{\rm{S}}{\bf{s}}_k^{\rm{H}}
\end{split}
\end{equation}
\vspace{-5pt}where ${{\bf R}_{{{\bf{g}}_{vi}}}} = \mathbb{E} \left[ {{\bf{g}}_{vi}^{\rm{H}}{{\bf{g}}_{vi}}} \right]$, ${{\bf{R}}_{{{\bf{N}}_v}}} = \mathbb{E}\left[ {{\bf{N}}_v^{\rm{H}}{{\bf{N}}_v}} \right]$ are the channel and noise correlation matrices.

\subsection{Stage 2: Uplink Data Transmission}
In the second stage, UEs send UL data to their associated UL BS and, the data of decoupled UEs are first decoded separately at each SBS with the estimated channels and then sent to the MBS via backhaul. However, the desired UL data signal at the SBSs is subject to high co-channel interference, and could hardly be decoded correctly using linear detectors. In this paper, discrete symbols are used to transmit data to combat severe interference. We assume that a UE transmits binary phase-shift-keying (BPSK)-coded UL data of totally $\tau_{\rm{D}}$ symbols to its associated UL BS. The matrix ${\bf{X}} = {\left[ {{\bf{x}}_1^{\rm{T}}, \cdots {\bf{x}}_k^{\rm{T}}, \cdots ,{\bf{x}}_K^{\rm{T}}} \right]^{\rm{T}}}$ is used to represent data sequences of UEs with each element randomly chosen from the set $\left\{ P_{\rm{D}}, -P_{\rm{D}} \right\}$ and ${\bf{x}}_k \in {\mathbb{C}^{1 \times \tau_{\rm{D}} }}$ is the UL data for the $k$th UE. Three linear receivers, maximal-ratio combining (MRC), zero-forcing (ZF) and MMSE are considered separately in the process of data decoding.

Firstly, the received signal at the $v$th SBS is expressed as
  \begin{equation}\label{}
    {{{\bf{Y}}}_v^{\rm{D}}} =  {{\bf{G}}_v}{\bf{X}} + {{{\bf{\tilde N}}}_v} = \sum\limits_{k = 1}^K {{{\bf{g}}_{vk}}{{\bf{x}}_k}}  + {{{\bf{\tilde N}}}_v},
  \end{equation}
where ${{{\bf{\tilde N}}}_v} \in\mathbb{C}^{N \times \tau_{\rm{D}}}$ is AWGN with each element subject to $\mathcal{CN}(0,N_{0})$ independently.

\subsubsection{ZF and MRC Detectors}
We use the estimated channel in Stage 1 to recover the UL data of the $k$th UE and, assume that the $k$th UE is associated with the $v$th SBS in UL. The MRC detector only needs CSI of the $k$th UE, while in this scheme the SBS has the CSI knowledge of UEs served by itself to perform ZF detection. Therefore, a new channel matrix is defined as ${{\mathcal {G}}_{vk}} = [{{{\bf{\hat g}}}_{vk}}, \cdots ]$, which involves the estimated channels of UEs associated with the $v$th SBS, knowing that the $k$th UE is included at least. Thus, the combination matrix ${\bf{A}}$ for MRC and ZF detectors are given by
  \begin{equation}\label{ZFMRC}
  {\bf{A}} =
  \begin{cases}
 {{{\bf{\hat g}}}_{vk}}{\left( {{\bf{\hat g}}_{vk}^{\rm{H}}{{{\bf{\hat g}}}_{vk}}} \right)^{ - 1}}, & {\mbox{for MRC}},\\
  {{\mathcal G}_{vk}}{\left( {{\mathcal G}_{vk}^{\rm{H}}{{\mathcal G}_{vk}}} \right)^{ - 1}},& {\mbox{for ZF}}.
  \end{cases}
  \end{equation}

Then the estimated data sequence of the $k$th UE after applying the combination matrix becomes
\begin{equation}\label{}
{{{\bf{\hat x}}}_k^{\rm ZF,MRC}} =  {\mathop{\rm Decoder}\nolimits} \left( {\left\lceil {{{\bf{A}}^{\rm{H}}}{{{\bf{Y}}}_v^{\rm{D}}}} \right\rceil _1} \right),
\end{equation}
where ${\left\lceil {\bf{B}} \right\rceil _i}$ performs the operation of taking the $i$th row of ${\bf{B}}$ and ${\rm Decoder}(\cdot)$ is the conventional BPSK decoder based on the maximum a posteriori probability (MAP) criterion. Note that in (\ref{ZFMRC}), the formation of combination matrix for ZF and MRC are similar. If an SBS only serves one UE, the matrix for ZF will degrade to that for MRC, which is the reason we take two detectors into consideration together.

\subsubsection{MMSE Detector}
In order to achieve better BER performance, MMSE detector is applied before decoding at each SBS. For MMSE detector with imperfect CSI, the standard way is to define MMSE channel estimation error as ${\bf{\tilde G}}_v = {{\bf{G}}_v} - {{{\bf{\hat G}}}_v}$, and the received signal can thus be rewritten as
\begin{equation}\label{MMSEStage2}
    {{{\bf{ Y}}}_v^{\rm{D}}} = {{{\bf{\hat G}}}_v}{\bf{X}} + {{{\bf{\tilde G}}}_v}{\bf{X}} + {{{\bf{\tilde N}}}_v}.
  \end{equation}

In what follows, we can treat the estimated channel as the real channel and the estimation error as independent noise, with the statistical property of MMSE \cite{steven1993fundamentals}. Therefore, the MMSE combination matrix for the $k$th UE can be written as
\begin{equation}\label{CombMMSE}
 {\bf{C}}_{vk}^{{\rm{UD,opt}}}
 %= {\bf{\hat g}}_{vk}^{\rm{H}}{\left( {{{{\bf{\hat G}}}_v}{\bf{\hat G}}_v^{\rm{H}} + \frac{1}{{{P_{\rm{D}}}}}\mathbb{E}\left[ {{{{\bf{\tilde G}}}_v}{\bf{X}}{{\bf{X}}^{\rm{H}}}{\bf{\tilde G}}{{_v^{\rm{H}}}}} \right] + \frac{1}{{{P_{\rm{D}}}}}\mathbb{E}\left[ {{{{\bf{\tilde N}}}_v}{\bf{\tilde N}}_v^{\rm{H}}} \right]} \right)^{ - 1}}\\
 ={\bf{\hat g}}_{vk}^{\rm{H}}{\left( {{{{\bf{\hat G}}}_v}{\bf{\hat G}}_v^{\rm{H}} + \left( {\sum\limits_{k = 1}^K {\frac{{{N_0}\beta _{vk}^{\rm{S}}}}{{{N_0} + \beta _{vk}^{\rm{S}}{P_{\rm{T}}}{\tau _{\rm{T}}}}}}  + \frac{{{N_0}}}{{{P_{\rm{D}}}}}} \right){\bf{I}}} \right)^{ - 1}}.
\end{equation}

After multiplied by the combination matrix ${\bf{C}}_{vk}^{{\rm{UD,opt}}}$, the estimated signal is sent to the BPSK decoder and the decoded data for the $k$th UE can be expressed as
\begin{equation}\label{}
{\bf{\hat x}}_k^{{\rm{MMSE}}} = {\mathop{\rm Decoder}\nolimits} \left( {{\bf{C}}_{vk}^{{\rm{UD,opt}}}{{{\bf{ Y}}}_v^{\rm{D}}}} \right).
\end{equation}\vspace{-20pt}

Here, three detectors are considered to recover the UEs' UL data at each SBS and the decoded data is then transferred to the MBS via backhaul for the data-aided channel estimation scheme.

\subsection{Stage 3: Data-Aided Channel Estimation}
As described in the frame structure, during the last stage, the MBS needs not only to proceed with the UL channel estimation and UL data detection for its own UEs but also to listen and record both UL training and data signal. As a result, in this stage, we can utilize the recorded UL signal along with the known UL training sequences and the decoded UL data to perform channel estimation at the MBS\footnote{When channel varies fast, it can be estimated with pilot and part of decoded UL data, herein decoded data is fully utilized.}. With UL data signal and decoded UL data sequences, there is more information for the MBS to improve the estimated channel accuracy of decoupled UEs. First, the received signal at the MBS during the first two stages can be jointly expressed as
\begin{equation}\label{MBS_Rcv}
  {\bf{M}^{\rm{PD}}} = {\bf{HW}} + {{\bf{Z}}} = \sum\limits_{k = 1}^K {{{\bf{h}}_k}{{\bf{w}}_k}}  + {{\bf{Z}}},
\end{equation}
where ${\bf{W}} = \left[ {{\bf{S}},{\bf{X}}} \right]$ with ${{{\bf{w}}_k}}$ being its $k$th row and ${{\bf{Z}}} \in \mathbb{C}^{M\times{(\tau_{\rm{T}}+\tau_{\rm{D}})}}$ is the AWGN noise at the MBS with each element an i.i.d.~$\mathcal{CN}(0,N_0)$ random variable.

Next, the MMSE channel estimator is adopted to recover the channels of decoupled UEs, by utilizing the UL training sequences and the decoded UL data sequences. Hence, the estimated channel from the $k$th UE to MBS can be written as
\begin{equation}\label{}
   {\bf{\hat h}}_{k}^{{\rm{MMSE}}} = {{\bf{M}^{\rm{PD}}}}{\bf{C}}_{k}^{{\rm{DA,opt}}},
\end{equation}
where
\begin{equation}\label{Gen_DA_MMSE}
{\mathbf{C}}_{vk}^{{\text{DA,opt}}}  = \left[ {\sum\limits_{i = 1}^K {\beta _k^{\text{M}} \mathbb{E}\left[ {{\mathbf{w}}_k^{\text{H}} {\mathbf{w}}_k } \right] + N_0 \mathbb{E}\left[ {{\mathbf{Z}}^{\text{H}} {\mathbf{Z}}} \right]} } \right]^{ - 1} \beta _k^{\text{M}} \mathbb{E}\left[ {{\mathbf{w}}_k^{\text{H}} } \right],
\end{equation}
thus, the channels from the decoupled UEs to the DL BS are obtained by this data-aided method.

Note that we only give the general expression of the combination matrix for the data-aided method in (\ref{Gen_DA_MMSE}). In the error-free scenario, we can easily obtain the result of $\mathbb{E}\left[ {{\mathbf{w}}_k^{\text{H}} {\mathbf{w}}_k } \right]$ in (\ref{Gen_DA_MMSE}) from the prior information about the UL data sequences, however, the random errors in the decoded data sequences are non-negligible in a more practical scenario. Therefore, the problems of how to estimate the BER performance of decoded data and how to model the random errors in the decoded sequences will be discussed in the following section. Moreover, the specific combination matrix for the data-aided scheme and channel estimation performance of different estimators will also be analyzed.

\section{Performance Analysis}
In this section, we conduct performance analysis to compare the data-aided scheme with conventional channel estimation techniques such as MMSE and LS. First of all, to evaluate the performance of different channel estimation methods, NMSE is adopted as the performance metric, which is defined as
\begin{equation}\label{NMSE_Criterio}
{\rm{NMSE}}{\rm{ = }}10{\log _{10}}\left( {\frac{{\mathbb{E}\left[ {{{\left\| {{{\bf{g}}} - {{{\bf{\hat g}}}}} \right\|}^2}} \right]}}{\mathbb{E}{\left[ {{{\left\| {{{\bf{g}}}} \right\|}^2}} \right]}}} \right)
\quad\left( {{\mbox{in dB}}} \right),
\end{equation}
where ${\bf{g}}$ represents the real channel vector and ${{\bf{\hat g}}}$ is the estimated one.

\subsection{Performance of Conventional Channel Estimation Methods}
In this subsection, we use the above metric to analyze the NMSE performances of conventional channel estimation methods by only taking the UL training sequences into account. Herein, we take NMSE of the $k$th UE associated with the $v$th SBS in Stage 1 as an example. Although the NMSE expressions here are based on the channel between a UE and an SBS, they can be easily generalized to the case of a UE connected to the MBS, which will be regarded as the performance benchmarks of conventional methods.

Now, we give the performance of the LS channel estimator. From (\ref{EstChSC}) and the NMSE definition in (\ref{NMSE_Criterio}),
the NMSE for the LS estimator of the $k$th UE associated with the $v$th SBS is
\begin{equation}\label{LSNMSE}
 {\mathcal J}_{vk}^{{\rm{S,LS}}} = 10{\log _{10}}\left( {\frac{{{N_0}}}{{\beta _{vk}^{\rm{S}}{\tau_{\rm{T}}}{P_{\rm{T}}}}}} \right).
\end{equation}

%the numerator of ${\rm{NMSE}}$ for LS can be calculated as
%\begin{equation}\label{}
%\mathbb{E}\left[{{{\left\| {{{\bf{g}}_{vk}} - {\bf{\hat g}}_{vk}^{{\rm{LS}}}} \right\|}^2}} \right]
%= \mathbb{E}\left[ {{{\left\| {\frac{1}{{{\tau_{\rm{T}}}{P_{\rm{T}}}}}{{\bf{N}}_v}{\bf{s}}_k^{\rm{H}}} \right\|}^2}} \right]\\
%%= \frac{1}{{P_{\rm{T}}^2}{\tau_{\rm{T}}^2}}{{\bf{s}}_k}\mathbb{E}\left[ {{\bf{N}}_v^{\rm{H}}{{\bf{N}}_v}} \right]{\bf{s}}_k^{\rm{H}}\\
%%= \frac{1}{{P_{\rm{T}}^2}{\tau_{\rm{T}}^2}}{{\bf{s}}_k}N{N_0}{{\bf{I}}_{{\tau _{\rm{T}}}}}{\bf{s}}_k^{\rm{H}}\\
%= \frac{{N{N_0}}}{{{\tau_{\rm{T}}}{P_{\rm{T}}}}},
%\end{equation}
%and the denominator can be straightforwardly written as
%\begin{equation}\label{ExpTrueCh}
%\mathbb{E}\left[ {{{\left\| {{{\bf{g}}_{vk}}} \right\|}^2}} \right]
%= \mathbb{E}\left[ {\beta _{vk}^{\rm{S}}{{\left( {{\bf{g}}_{sk}^{\rm{W}}} \right)}^{\rm{H}}}{\bf{g}}_{sk}^{\rm{W}}} \right]
%= N {\beta _{vk}^{\rm{S}}}.
%\end{equation}
\vspace{-5pt}
\begin{remark}
From the above expression, it is observed that NMSE for the conventional LS estimator is related to ${N_0}, \beta _{vk}^{\rm{S}}, {\tau_{\rm{T}}}$ and ${P_{\rm{T}}}$. It states that once a UE is located and the BS is stationary, larger pilot power and longer pilot sequence are two means to improve channel estimation accuracy.
\end{remark}

Before analyzing the NMSE for the MMSE estimator, we rewrite the estimated channel of the $k$th UE to the $v$th SBS as\vspace{-5pt}
 \begin{equation}\label{VarEstCh}
    \begin{split}
{{{\bf{\hat g}}}_{vk}}
%&= \left( {{{\bf{G}}_v}{\bf{S}} + {{\bf{N}}_v}} \right){\left( {{{\bf{S}}^{\rm{H}}}{{\bf{R}}_{\bf{g}}}{\bf{S}} + {{\bf{R}}_{{{\bf{N}}_v}}}} \right)^{ - 1}}{\bf{s}}_k^{\rm{H}}{{\bf R}_{{{\bf{g}}_{vk}}}}\\
% &\overset{(a)}{=} {R_{{{\bf{g}}_{vk}}}}\left( {{{\bf{G}}_v}{\bf{S}} + {{\bf{N}}_v}} \right)\left\{ {\frac{1}{{N{N_0}}}{\bf{s}}_k^{\rm{H}} - \frac{{{\tau _{\rm{T}}}{P_{\rm{T}}}}}{{{N^2}N_0^2}}{{\bf{S}}^{\rm{H}}}{{\left( {{\bf{R}}_{\bf{g}}^{ - 1} + \frac{{{\tau _{\rm{T}}}{P_{\rm{T}}}}}{{N{N_0}}}{\bf{I}}} \right)}^{ - 1}}{\bf{Ss}}_k^{\rm{H}}} \right\} \\
% &= \frac{{{R_{{{\bf{g}}_{vk}}}}}}{{\left( {N{N_0} + {R_{{{\bf{g}}_{vk}}}}{P_{\rm{T}}}{\tau _{\rm{T}}}} \right)}}\left( {{{\bf{G}}_v}{\bf{Ss}}_k^{\rm{H}} + {{\bf{N}}_v}{\bf{s}}_k^{\rm{H}}} \right)\\
 &= \frac{{{{\bf R}_{{{\bf{g}}_{vk}}}}}}{{\left( {N{N_0} + {{\bf R}_{{{\bf{g}}_{vk}}}}{P_{\rm{T}}}{\tau _{\rm{T}}}} \right)}}\left( {{\tau _{\rm{T}}}{P_{\rm{T}}}{{\bf{g}}_{vk}} + {{\bf{N}}_v}{\bf{s}}_k^{\rm{H}}} \right),
    \end{split}
 \end{equation}
\vspace{-5pt}where ${{\bf{R}}_{\bf{g}}} = \mathbb{E} \left[ {{\bf{G}}_v^{\rm{H}}{{\bf{G}}_v}} \right] ={\rm{diag}}\left( {{{\bf R}_{{{\bf{g}}_{v1}}}} \dots{{\bf R}_{{{\bf{g}}_{vi}}}} \dots {{\bf R}_{{{\bf{g}}_{vK}}}}} \right)$, ${{\bf R}_{{{\bf{g}}_{vi}}}} = \mathbb{E} \left[ {{\bf{g}}_{vi}^{\rm{H}}{{\bf{g}}_{vi}}} \right]$.
%  \begin{equation}\label{inequ_indenity1}
%  {\left( {{\bf{A + UBV}}} \right)^{ - 1}} = {{\bf{A}}^{ - 1}} - {{\bf{A}}^{ - 1}}{\bf{U}}{\left( {{{\bf{B}}^{ - 1}} + {\bf{V}}{{\bf{A}}^{ - 1}}{\bf{U}}} \right)^{ - 1}}{\bf{V}}{{\bf{A}}^{ - 1}}.
%\end{equation}

%Since the training sequences satisfy ${\bf{S}}{{\bf{S}}^{\rm{H}}} = \tau_{\rm{T}}{{P_{\rm{T}}}}{{\bf{I}}_K}$ and each element of ${{\bf{N}}_v}$ is i.i.d.~$\mathcal{CN}(0,N_0)$, it can be inferred that the matrix ${{{\bf{N}}_v}{\bf{s}}_k^{\rm{H}}}$ is distributed as $\mathcal{CN}(0,{\tau _{\rm{T}}}{P_{\rm{T}}}N_0{\bf{I}})$.
Observing the equation (\ref{VarEstCh}), ${{{\bf{g}}_{vk}}}$ and ${{{\bf{N}}_v}{\bf{s}}_k^{\rm{H}}}$ are independent complex Gaussian random variables, therefore, it is the property of Gaussian distribution that ${{{\bf{\hat g}}}_{vk}}$ is also a complex Gaussian variable following
\begin{equation}\label{}
{{{\bf{\hat g}}}_{vk}} \sim \mathcal{CN}\left( {0,\frac{{{{\left( {\beta _{vk}^{\rm{S}}} \right)}^2}{P_{\rm{T}}}{\tau _{\rm{T}}}}}{{{N_0} + \beta _{vk}^{\rm{S}}{P_{\rm{T}}}{\tau _{\rm{T}}}}}{\bf{I}}} \right).
\end{equation}
\vspace{-5pt}
Then we denote the channel estimation error of the MMSE estimator as ${{{\bf{\tilde g}}}_{vk}} \buildrel \Delta \over = {{\bf{g}}_{vk}} - {{{\bf{\hat g}}}_{vk}}$, with the help of the orthogonal property of the MMSE estimator, ${{{\bf{\tilde g}}}_{vk}}$ is independent of ${{{\bf{\hat g}}}_{vk}}$. Thus, we directly obtain the channel estimation error vector, which is distributed as
\begin{equation}\label{EstChErr}
{{{\bf{\tilde g}}}_{vk}} \sim \mathcal{CN}\left( {0,\frac{{{N_0}\beta _{vk}^{\rm{S}}}}{{{N_0} + \beta _{vk}^{\rm{S}}{P_{\rm{T}}}{\tau _{\rm{T}}}}}{\bf{I}}} \right).
\end{equation}

In parallel to the LS estimator, the NMSE for the MMSE estimator of the $k$th UE associated with the $s$th SBS can be calculated by using (\ref{EstChErr})
\begin{equation}\label{MMSENMSE}
{\cal J}_{vk}^{{\rm{S}},{\rm{MMSE}}}
%= 10{\log _{10}}\left( {\frac{\mathbb{E}{\left[ {{{\left\| {{{\bf{g}}_{vk}} - {{{\bf{\hat g}}}_{vk}}} \right\|}^2}} \right]}}{\mathbb{E}{\left[ {{{\left\| {{{\bf{g}}_{vk}}} \right\|}^2}} \right]}}} \right)
%&= 10{\log _{10}}\left( {\frac{{\frac{{N{N_0}\beta _{vk}^{\rm{S}}}}{{{N_0} + \beta _{vk}^{\rm{S}}{P_{\rm{T}}}{\tau _{\rm{T}}}}}}}{{N\beta _{vk}^{\rm{S}}}}} \right)\\
= 10{\log _{10}}\left( {\frac{{{N_0}}}{{{N_0} + \beta _{vk}^{\rm{S}}{P_{\rm{T}}}{\tau _{\rm{T}}}}}} \right).
\end{equation}

\begin{remark}
Similar to the LS estimator, NMSE for the MMSE estimator is a function of ${\tau _{\rm{T}}}$, ${P_{\rm{T}}}$, $\beta _{vk}^{\rm{S}}$ and $N_0$. We can also increase pilot length and transmitted power to improve channel estimation performance once the locations of UE and SBS are both set. Comparing two NMSE expressions, the only difference appears to be the extra noise variance term $N_0$ in the numerator of NMSE for MMSE. Hence, the performance gain of the MMSE estimator over LS is obtained in the regime of low SNR and drops to $0$ when SNR gets higher.
\end{remark}

\subsection{Performance of Data-Aided Channel Estimation Method}
To evaluate our proposed data-aided method, the process of UL data transmission is analyzed and the BER is obtained since the decoded data is used to aid channel estimation and may highly affect channel estimation in stage 3. Here, we assume that the CSI acquired by the MMSE estimator is applied and the MMSE detector is performed for data decoding in Stage 2, as MMSE detector performs better in interference-limited systems.
We first consider the BER performance of UL data in Stage 2 and the well-know BER expression for binary modulations is given by\cite{li2006distribution}
%\begin{equation}\label{OriBER}
%{\mathop{\rm BER}\nolimits} \left( {{\rm{SINR}}} \right) = \int_{\sqrt {{\rm{SINR}}} }^\infty  {\frac{1}{{\sqrt {2\pi } }}\exp \left( { - \frac{1}{2}{t^2}} \right){\rm{d}}t}.
%\end{equation}
%
%Considering the randomness of $\rm SINR$ (signal-to-interference plus noise ratio), the ergodic BER for the MMSE detector can be further written as
\begin{equation}\label{DEFBERALL}
{\mathop{\rm BER}\nolimits}  = \int_0^\infty  {{f_{{\rm{SINR}}}}\left( x \right)\int_{\sqrt x }^\infty  {\frac{1}{{\sqrt {2\pi } }}\exp \left( { - \frac{1}{2}{t^2}} \right){\rm{d}}t}~{\rm{d}}x} ,
\end{equation}
where ${{f_{{\rm{SINR}}}}\left( x \right)}$ is the probability density function (p.d.f.) of $\rm SINR$ (signal-to-interference plus noise ratio).

As a matter of fact, the exact distribution of $\rm SINR$ for the MMSE detector has been derived in \cite{gao1998theoretical}, but the distribution is too complex to obtain a closed-form expression of BER. Therefore, we resort to the well-known tractable Gamma distribution, whose p.d.f.~is given as
\begin{equation}\label{}
{f_{{\rm{Gamma}}}}\left( {x;\alpha ,\xi } \right) = \frac{{{x^{\alpha  - 1}}{e^{ - \frac{x}{\xi }}}}}{{\Gamma \left( \alpha  \right){\xi ^\alpha }}}
\end{equation}
\vspace{-5pt}to approximate the result.

Similar approach to obtain closed-form solutions has been applied in many previous efforts, see, e.g., \cite{li2006distribution,ngo2013energy}. Following the similar procedure in \cite{li2006distribution}, by determining two parameters of the approximated Gamma distribution, a closed-form expression of BER can be obtained. However, different from \cite{li2006distribution,ngo2013energy}, in our case, the antenna number at each SBS is assumed less than the number of UEs. In the following, two parameters of Gamma distribution are determined by moment matching.

We denote ${{\rm{SIN}}{{\rm{R}}_{vk}}}$ as the $\rm SINR$ of the $k$th UE served by the $v$th SBS with the MMSE detector. From (\ref{MMSEStage2})(\ref{CombMMSE}), ${{\rm{SINR}}{_{vk}}}$ can be expressed as
\begin{equation}\label{}
{\rm{SIN}}{{\rm{R}}_{vk}} = \frac{1}{{{{\left( {{{\left( {{\bf{I}} + {\rho _v}{\bf{\hat G}}_v^{\rm{H}}{{{\bf{\hat G}}}_v}} \right)}^{ - 1}}} \right)}_{kk}}}} - 1,
\end{equation}
where ${\rho _v} = {\left( {\sum\limits_{k = 1}^K {\frac{{{N_0}\beta _{vk}^{\rm{S}}}}{{{N_0} + \beta _{vk}^{\rm{S}}{P_{\rm{T}}}{\tau _{\rm{T}}}}}}  + \frac{{{N_0}}}{{{P_{\rm{D}}}}}} \right)^{ - 1}}$ and ${\left(  \cdot  \right)_{kk}}$ represents the $(k,k)$th element of a matrix.

%\begin{lemma}
%For a square matrix ${\bm{\Omega }} = \left( {{{\bm{\omega }}_1},{{\bm{\omega }}_2}, \cdots ,{{\bm{\omega }}_k}, \cdots ,{{\bm{\omega }}_K}} \right) $, then
%\begin{equation*}
%{\left( {{{\bm{\Omega }}^{ - 1}}} \right)_{kk}} = {\left( {{{\bm{\omega }}_{kk}} - {\bm{\omega }}_{k\left( { - k} \right)}^{\rm{H}}{{\left( {{{\bm{\omega }}_{\left( { - k, - k} \right)}}} \right)}^{ - 1}}{{\bm{\omega }}_{k\left( { - k} \right)}}} \right)^{ - 1}},
%\end{equation*}
%where ${{{\bm{\omega }}_{k\left( { - k} \right)}}}$ is the $k$th column of ${\bf{\Omega }}$ with $k$th element removed and ${{{\bm{\omega }}_{\left( { - k, - k} \right)}}}$ is the matrix with $k$th column and $k$th row removed from ${\bf{\Omega }}$.
%\begin{proof}
%\end{proof}
%\end{lemma}

Then by applying \cite[(8)]{li2006distribution}, ${{\rm{SIN}}{{\rm{R}}_{vk}}}$ can be further expressed as
\begin{equation}\label{SINRDecom}
{\rm{SIN}}{{\rm{R}}_{vk}}
%&= \frac{1}{{{{\left( {{{\left( {{\bf{I}} + {\rho _v}{\bf{\hat G}}_v^{\rm{H}}{{{\bf{\hat G}}}_v}} \right)}^{ - 1}}} \right)}_{kk}}}} - 1 \\
%&= \rho_v {\bf{\hat g}}_{vk}^{\rm{H}}{{{\bf{\hat g}}}_{vk}} - {\rho_v ^2}{\bf{\hat g}}_{vk}^{\rm{H}}{{{\bf{\hat G}}}_{v\left( { - k} \right)}}{\left( {{\bf{I}} + {\rho _v}{\bf{\hat G}}_{v\left( { - k} \right)}^{\rm{H}}{{{\bf{\hat G}}}_{v\left( { - k} \right)}}} \right)^{ - 1}}{\bf{\hat G}}_{v\left( { - k} \right)}^{\rm{H}}{{{\bf{\hat g}}}_{vk}} \\
={\rho _v}{\bf{\hat g}}_{vk}^{\rm{H}}{{{\bf{\hat g}}}_{vk}} -
\rho _v^2{\bf{\hat g}}_{vk}^{\rm{H}}{{{\bf{\hat G}}}_{v\left( { - k} \right)}}{\bf{\hat G}}_{v\left( { - k} \right)}^{\rm{H}}{\left( {{\bf{I}} + {\rho _v}{{{\bf{\hat G}}}_{v\left( { - k} \right)}}{\bf{\hat G}}_{v\left( { - k} \right)}^{\rm{H}}} \right)^{ - 1}}{{{\bf{\hat g}}}_{vk}},
\end{equation}
\vspace{-5pt}where ${{{\bf{\hat G}}}_{v\left( { - k} \right)}}$ is the matrix ${{{\bf{\hat G}}}_{v}}$ with the $k$th column removed and ${{{\bf{\hat g}}}_{vk}}$ is the $k$th column of ${{{\bf{\hat G}}}_{v}}$.

Considering the singular value decomposition (SVD), we have ${{{\bf{\hat G}}}_{v\left( { - k} \right)}} = {\bf{UD}}{{\bf{V}}^{\rm{H}}}$, ${\bf{U}} \in {\mathbb{C}^{N \times N}},{\bf{D}} \in {\mathbb{C}^{N \times N}},{{\bf{V}}^{\rm{H}}} \in {\mathbb{C}^{N \times \left( {K - 1} \right)}}$, ${\bf{U}}{{\bf{U}}^{\rm{H}}} = {{\bf{U}}^{\rm{H}}}{\bf{U}} = {{\bf{I}}_N}$ and ${{\bf{V}}^{\rm{H}}}{\bf{V}} = {{\bf{I}}_N}$. Then applying SVD to  (\ref{SINRDecom}), we get\vspace{-5pt}
\begin{align}\label{SINRSplit}
{\rm{SIN}}{{\rm{R}}_{vk}} &= {\rho _v}{\bf{\hat g}}_{vk}^{\rm{H}}{{{\bf{\hat g}}}_{vk}} - \rho _v^2{\bf{\hat g}}_{vk}^{\rm{H}}{\bf{U}}{{\bf{D}}^2}{\left( {{\bf{I}} + {\rho _v}{{\bf{D}}^2}} \right)^{ - 1}}{{\bf{U}}^{\rm{H}}}{{{\bf{\hat g}}}_{vk}} \notag\\
&= {\rho _v}{{\bm{\varphi }}^{\rm{H}}}{\bm{\varphi }} - \rho _v^2{{\bm{\varphi }}^{\rm{H}}}{{\bf{D}}^2}{\left( {{\bf{I}} + {\rho _v}{{\bf{D}}^2}} \right)^{ - 1}}{\bm{\varphi }} \notag\\
&= {\rho _v}\sum\limits_{i = 1}^N {\frac{{{{\left\| {{\varphi _i}} \right\|}^2}}}{{1 + {\rho _v}d_i^2}}},
\end{align}
\vspace{-5pt}where ${{\bf{U}}^{\rm{H}}}{{{\bf{\hat g}}}_{vk}} \buildrel \Delta \over = {\bm{\varphi }}$ and ${d_i}$ is the $i$th diagonal element of ${\bf{D}}$. As ${\bf{U}}$ is a unitary matrix, it can be proved that ${\bm{\varphi }}$ has the same distribution as ${{{\bf{\hat g}}}_{vk}}$.

Next, we proceed our analysis with the help of some known results of the empirical eigenvalue distribution of the product of two random matrices \cite{silverstein1995strong,bai2004clt}. Although these results are obtained under the limiting condition, it is shown in \cite{li2006distribution} that this approximation is, to some extent, accurate even for very small dimensions. In our case, those results are extended to the scenario where the number of antennas is less than the number of UEs, namely, $N<K$.

The empirical eigenvalue distribution (ESD) of ${\rho _v}{{{\bf{\hat G}}}_{v\left( { - k} \right)}}{\bf{\hat G}}_{v\left( { - k} \right)}^{\rm{H}}$, denoted by ${{\bf{\hat J}}}$, converges to a measure ${\bf{J}}$, whose Stieltjes transform, denoted by ${\cal T}\left( z \right)$, is given as\vspace{-5pt}
\begin{equation}\label{}
{\cal T}\left( z \right) \buildrel \Delta \over = \int {\frac{1}{{x - z}}{\bf{J}}\left( {{\rm{d}}x} \right)}.
\end{equation}
\vspace{-5pt}
According to the results in \cite{hoydis2010asymptotic}, this integral can be approximated by\vspace{-5pt}
\begin{equation}\label{Iter1}
{\cal T}\left( z \right) \approx {\left( {\sum\limits_{i \ne k}^K {\frac{{{{\bf{V}}_i}}}{{1 + {\mathop{\rm Tr}\nolimits} \left( {{{\bf{V}}_i}{\cal T}\left( z \right)} \right)}}}  - z{\bf{I}}} \right)^{ - 1}},
\end{equation}
\vspace{-5pt}where ${{\bf{V}}_i} = {\rho _v}\hat \beta _{vk}^{\rm{S}}{{\bf{I}}_N}$ and $\hat \beta _{vk}^{\rm{S}} = \frac{{{{\left( {\beta _{vk}^{\rm{S}}} \right)}^2}{P_{\rm{T}}}{\tau _{\rm{T}}}}}{{{N_0} + \beta _{vk}^{\rm{S}}{P_{\rm{T}}}{\tau _{\rm{T}}}}}$. Similarly, the first derivative of ${\cal T}$ is found as\vspace{-5pt}
\begin{equation}\label{Iter2}
{\cal T}'\left( z \right) \buildrel \Delta \over =\int {\frac{1}{{{{\left( {x - z} \right)}^2}}}{\bf{J}}\left( {{\rm{d}}x} \right)}  = {{\cal T}^2}\left( z \right) \\\approx {\left( {\sum\limits_{i \ne k}^K {\frac{{{{\bf{V}}_i}}}{{1 + {\mathop{\rm Tr}\nolimits} \left( {{{\bf{V}}_i}{\cal T}\left( z \right)} \right)}}}  - z{\bf{I}}} \right)^{ - 2}}.
\end{equation}

Note that ${\cal T}\left( z \right)$ and ${\cal T}'\left( z \right)$ are well defined at $z=-1$ by the bounded convergence theorem\cite{bai2010spectral,durrett2010probability}. Then, we have
\begin{equation}\label{DefESINR}
\frac{{{\mathop{\rm Tr}\nolimits} \left( {\bf{\Lambda }} \right)}}{N} = \frac{1}{N}\sum\limits_{i = 1}^N {\frac{1}{{1 + {\rho _v}d_i^2}}}  = \int {\frac{1}{{x + 1}}{\bf{\hat J}}\left( {{\rm{d}}x} \right)}
\xrightarrow{~p~} \int {\frac{1}{{x + 1}}{\bf{J}}\left( {{\rm{d}}x} \right)}  = {\cal T}\left( { - 1} \right) \buildrel \Delta \over = \mu,
\end{equation}
\begin{equation}\label{}
\frac{{{\mathop{\rm Tr}\nolimits} \left( {{{\bf{\Lambda }}^2}} \right)}}{N} = \frac{1}{N}\sum\limits_{i = 1}^N {\frac{1}{{{{\left( {1 + {\rho _v}d_i^2} \right)}^2}}}}  = \int {\frac{1}{{{{\left( {x + 1} \right)}^2}}}{\bf{\hat J}}\left( {{\rm{d}}x} \right)} 
\xrightarrow{~p~} \int {\frac{1}{{{{\left( {x + 1} \right)}^2}}}{\bf{J}}\left( {{\rm{d}}x} \right)}  = {\cal T}'\left( { - 1} \right) \buildrel \Delta \over = {\sigma ^2},
\end{equation}
where $\xrightarrow{~p~}$ means convergence in probability and ${\bf{\Lambda }} \buildrel \Delta \over =  {\rm{Diag}}\left( {{\lambda _1}, \dots ,{\lambda _i}, \dots ,{\lambda _N}} \right),{\lambda _i} = \frac{1}{{1 + {\rho _v}d_i^2}}$.

By solving (\ref{Iter1}) and (\ref{Iter2}), we can obtain $\mu$ and ${\sigma ^2}$ and have the following lemma.

\begin{lemma}
The first two moments of ${{\rm{SIN}}{{\rm{R}}_{vk}}}$ can be approximated by
\begin{align}
\mathbb{E}\left[ {{\rm{SIN}}{{\rm{R}}_{vk}}} \right] &\approx N{\rho _v}\hat \beta _{vk}^{\rm{S}}\mu,\label{ESINRAPR}\\
{\mathop{\rm Var}\nolimits} \left[ {{\rm{SIN}}{{\rm{R}}_{vk}}} \right] &\approx N{\left( {{\rho _v}\hat \beta _{vk}^{\rm{S}}} \right)^2}{\sigma ^2}.\label{VARSINRAPR}
\end{align}
\begin{IEEEproof}
See Appendix A.
\end{IEEEproof}
\end{lemma}

Thus, with the first two moments of ${{\rm{SIN}}{{\rm{R}}_{vk}}}$, its p.d.f. is determined and the ergodic BER expression can be presented by the proposition below.

\begin{prop}
For the $k$th UE served by the $v$th SBS employing MMSE data detection with imperfect CSI (acquired by MMSE estimator), the BER of its UL data is expressed as
\begin{equation}\label{EstBER}
{{\mathop{\rm BER}\nolimits} _{vk}} = \frac{{\Gamma \left( {{\alpha _{vk}} + \frac{1}{2}} \right)}}{{\Gamma \left( {{\alpha _{vk}}} \right)2\sqrt {2\pi } }}\frac{{\xi _{vk}^{ - {\alpha _{vk}}}}}{{{\alpha _{vk}}{{\left( {\frac{1}{{{\xi _{vk}}}} + \frac{1}{2}} \right)}^{{\alpha _{vk}} + \frac{1}{2}}}}}
\times{}_2{F_1}\left( {1,{\alpha _{vk}} + \frac{1}{2};{\alpha _{vk}} + 1;\frac{{\frac{1}{{{\xi _{vk}}}}}}{{\frac{1}{{{\xi _{vk}}}} + \frac{1}{2}}}} \right),
\end{equation}
where we have ${\rm{SINR}}_{vk} \sim {\mathop{\rm Gamma}\nolimits} \left( {{\alpha _{vk}},{\xi _{vk}}} \right)$ with ${\alpha _{vk}} = N\frac{{{\mu ^2}}}{{{\sigma ^2}}},{\xi _{vk}} = {\rho _v}\hat \beta _{vk}^{\rm{S}}\frac{{{\sigma ^2}}}{\mu }$. $\Gamma \left(  \cdot  \right)$ represents gamma function and ${}_2F_1 \left(  \cdot  \right)$ is the hypergeometric function.

\begin{IEEEproof}
Based on (\ref{ESINRAPR}) and (\ref{VARSINRAPR}), the distribution of ${{\rm{SIN}}{{\rm{R}}_{vk}}}$ can be determined by moment matching. Then plugging its distribution in (\ref{DEFBERALL}), we obtain the final result after integration.
\end{IEEEproof}
\end{prop}

We note that both Gamma function and hypergeometric function are involved in the above result, which can be evaluated numerically but hardly shed any light on the BER performance. Hence, approximated expressions are derived in the following corollary.

\begin{corollary}
According to the result in (\ref{DEFBERALL}), we find that BER is a strict convex function. Thus, by applying Jensen's inequality, the ergodic BER can be lower bounded by
\begin{equation}\label{}
{\mathop{\rm BER}\nolimits} _{vk}^{{\rm{Lower}}} = {\mathop{\rm Q}\nolimits} \left( \sqrt{ \mathbb{E}{\left[ {{\rm{SIN}}{{\rm{R}}_{vk}}} \right]} } \right) = {\mathop{\rm Q}\nolimits} \left( {\sqrt {{\alpha _{vk}}{\xi _{vk}}} } \right),\vspace{-5pt}
\end{equation}
where ${\mathop{\rm Q}\nolimits} \left( x \right) = \int_x^\infty  {\frac{1}{{\sqrt {2\pi } }}\exp \left( { - \frac{1}{2}{t^2}} \right){\rm{d}}t}$. The lower bound of the ergodic BER is a monotonic decreasing function of the first moment of ${\rm{SIN}}{{\rm{R}}_{vk}}$, also of all the parameters of the first moment.
\end{corollary}

After UL data BER values are derived, we are now able to continue with the evaluation of the data-aided channel estimation, which utilizes MMSE estimation to recover the channels from the joint sequences consisting of known training sequences and decoded data with certain BER.

To do so, we model the joint sequences as follows. Recall the joint UL signal ${\mathbf{W}}$ in (\ref{MBS_Rcv}), which is the combination of known training sequences ${\mathbf{S}}$ and UL data sequences ${\mathbf{X}}$ of all UEs. Denote the block matrix of known training sequences and decoded data as ${\mathbf{\hat W}}=[{\mathbf{S}},{\mathbf{\hat X}}]$, and use ${\mathbf{E}}$ to represent the error matrix which can be defined as\vspace{-5pt}
\begin{equation}\label{}
{\mathbf{W}} \triangleq {\mathbf{\hat W}} \circ {\mathbf{E}},\vspace{-5pt}
\end{equation}
\vspace{-5pt}where $\circ$ represents Hadamard product of two matrices. Note that it can further be decomposed into two parts as ${\mathbf{\hat W}} \circ {\mathbf{E}} = \left[ {{\mathbf{S}},{\mathbf{\hat X}}} \right] \circ \left[ {{\mathbf{E}}_1, {\mathbf{E}}_2 } \right]$. As we know the training sequences exactly, ${{\mathbf{E}}_1 }$ is an all-one matrix, which represents an identity matrix in the Hadamard product, while ${{\mathbf{E}}_2 }$ indicates errors in the decoded sequences with all elements from the set $\left\{ {1, - 1} \right\}$. Then we can obtain the following statistical property of ${{\mathbf{E}}_2 }$ by utilizing the value of ${\text {BER}}$ obtained previously\vspace{-3pt}
\begin{equation}\label{}
\begin{aligned}
  &\mathbb{E}\left[ {e_{ij} } \right] = 1 - 2{\text{BER}}_{vi},\\
  &\mathbb{E}\left[ {\left\| {e_{ij} } \right\|^2 } \right] = 1,
\end{aligned}
\end{equation}
\vspace{-5pt}where $e_{ij}  = \left[ {{\mathbf{E}}_2 } \right]_{ij}$. Intuitively, we assume that each bit in ${\mathbf{\hat X}}$ from each UL data stream has the same probability of error, regardless of its location and what is actually sent, meaning that $e_{ij}$ for $j=1,\dots,\tau _{\text{D}}$ are i.i.d.~random variables, and the elements in ${\mathbf{E}}_2$ from different streams are independent. Therefore, the following proposition elucidates data-aided channel estimation with the MMSE estimator.

\begin{prop}
For the MMSE estimator, the recovered channel of the $k$th UE at the MBS by using data-aided channel estimation can be written as\vspace{-5pt}
\begin{align}
&{\mathbf{\hat h}}_k^{{\rm{MMSE}}}= \left( {{\mathbf{HW}} + {\mathbf{Z}}} \right){\mathbf{C}}_k^{{\rm{DA,opt}}},\\
&{\mathbf{C}}_k^{{\rm{DA,opt}}}= \left( {{\mathbf{P}} + N_0 {\mathbf{I}}} \right)^{ - 1} \left( {{\mathbf{\hat w}}_k^{\text{H}}  \circ \mathbb{E}\left[ {{\mathbf{e}}_k^{\text{H}} } \right]} \right)\beta _k^{\text{M}},
\end{align}
where ${\mathbf{e}}_k$ is the $k$th row of $\mathbf E$ with\vspace{-5pt}
\begin{equation}\label{}
\mathbb{E}\left [{\mathbf{e}}_k^{\text{H}} \right]= \left( {\begin{array}{*{20}c}
   {\underbrace {1 \cdots 1}_{\tau _{\text{T}} }} & {\underbrace {1 - 2{\text{BER}}_{vk}  \cdots 1 - 2{\text{BER}}_{vk} }_{\tau _{\text{D}} }}  \\

 \end{array} } \right)^{\text{H}},
\end{equation}
and ${\mathbf{P}} \triangleq {\mathbf{\hat P}} + \Delta {\mathbf{P}}$ with\vspace{-8pt}
\begin{equation}\label{}
{\mathbf{\hat P}} = \left( {\left[ {{\mathbf{S}},{\mathbf{\hat X}} \circ \mathbb{E}\left[ {{\mathbf{E}}_2 } \right]} \right]} \right)^{\text{H}} {\mathbf{R}}_{\mathbf{H}} \left( {\left[ {{\mathbf{S}},{\mathbf{\hat X}} \circ \mathbb{E}\left[ {{\mathbf{E}}_2 } \right]} \right]} \right),
\end{equation}
\begin{equation}\label{}
\Delta {\mathbf{P}} \triangleq \left( {\begin{array}{*{20}c}
   {\mathbf{0}} & {\mathbf{0}}  \\
   {\mathbf{0}} & \Delta S_{\mathbf{X}} {\mathbf{I}}_{\tau _{\text{D}} }
\end{array} } \right),
\end{equation}
where ${\mathbf{R}}_{\mathbf{H}}  = {\text{diag}}\left( {\beta _1^{\text{M}} , \dots ,\beta _k^{\text{M}} , \dots ,\beta _K^{\text{M}} } \right)$ and $\Delta S_{\mathbf{X}}  = P_{\text{D}} \sum\limits_{k=1}^K {\beta _k^{\text{M}} } \left\{ {1 -\left( {1 - 2{\text{BER}}_{vk} } \right)^2 } \right\}.$

The NMSE of the channel from the $k$th UE to the MBS by using the data-aided channel estimation method with MMSE can be calculated asymptotically as
\begin{equation}\label{}
\mathcal{J}_k^{{\rm{MMSE}}}  = 10\log _{10} \left( {\frac{1}
{{1 + \rho _k^{{\rm{DA}}} \beta _k^{\text{M}} }}} \right),
\end{equation}
where
\begin{equation}\label{}
\rho _k^{{\rm{DA}}}  = \frac{{\tau _{\text{T}} P_{\text{T}} }}
{{N_0 }} + \frac{{\tau _{\text{D}} P_{\text{D}} \left( {1 - 2{\text{BER}}_{vk} } \right)^2 }}
{{\Delta S_{\mathbf{X}}  + N_0 }}.
\end{equation}
\end{prop}

\begin{IEEEproof}
See Appendix B.
\end{IEEEproof}

\begin{remark}
From \emph{Proposition 2}, it is observed that the proposed data-aided channel estimation method introduces more parameters into NMSE, such as UL data transmission power, length of UL data sequence and BER of itself and other UEs. It is anticipated that better BER, higher data transmit power and longer UL data length will have positive effects on NMSE of data-aided estimation.
\end{remark}

\subsection{Comparison between Conventional and Data-Aided Channel Estimation}
Here, we first summarize the NMSE expressions derived previously in TABLE \ref{ComTable}. Note that results in the table describe NMSE performances of the estimated channel from decoupled UEs to the associated MBS with different channel estimators. The NMSE results in the first two columns are directly transformed from that of UEs associated with the SBSs in (\ref{LSNMSE}) and (\ref{MMSENMSE}).

\begin{table}[]
\centering
\caption{NMSE for different channel estimation methods}\label{ComTable}
\begin{tabular}{|c|c|c|}
\hline
\multicolumn{2}{|c|}{\bf Conventional}                                   & \multicolumn{1}{c|}{\textbf{\textbf{Data-Aided}}}                                 \\ \hline
\multicolumn{1}{|c|}{\textbf{\textbf{LS}}} & \multicolumn{1}{c|}{\textbf{\textbf{MMSE}}} & \textbf{\textbf{MMSE}}                                                      \\ \hline
\multirow{2}{*}{\begin{large}$10\log _{10} \left( {\frac{1}
{{\rho^{{\text{Con}}} \beta _k^{\text{M}} }}} \right)$\end{large}}                & \multirow{2}{*}{\begin{large}$10\log _{10} \left( {\frac{1}
{{1 + \rho^{{\text{Con}}} \beta _k^{\text{M}} }}} \right)$\end{large}}               & \multirow{2}{*}{\begin{large}$10\log _{10} \left( {\frac{1}
{{1 + \rho _k^{{\text{DA}}} \beta _k^{\text{M}} }}} \right)$\end{large}}                                               \\
                                  &                                  &                                                                  \\ \hline
\multicolumn{2}{|c|}{\begin{large}$\rho^{{\text{Con}}}  = \frac{{\tau _{\text{T}} P_{\text{T}} }}
{{N_0 }}$\end{large}}                                             & \begin{tabular}[c]{@{}l@{}}\begin{large}$
\rho _k^{{\text{DA}}}  = \frac{{\tau _{\text{T}} P_{\text{T}} }}
{{N_0 }} + \frac{{\tau _{\text{D}} P_{\text{D}} \left( {1 - 2{\text{BER}}_{vk} } \right)^2 }}{{\Delta S_{\mathbf{X}}  + N_0 }}$\end{large}\\\begin{small}$
\Delta S_{\mathbf{X}}  = P_{\text{D}} \sum\limits_k^K {\beta _k^{\text{M}} \left\{ {1 - \left( {1 - 2{\text{BER}}_{vk} } \right)^2 } \right\}}$\end{small}
\end{tabular} \\ \hline
\end{tabular}
\end{table}

In general, the table demonstrates that all NMSE results are parameterized by large-scale fading and a SNR-like term. Furthermore, after we combine these two parameters, it can be viewed as an effective signal power divided by noise power at the receiver, which corresponds to the meaning of an effective SNR. Thus, NMSE can be interpreted as the function of reverse effective SNR as a whole.

Next, we will examine the differences among these estimators and try to reveal some insights. To begin with, compared with the LS estimator, we can clearly observe that MMSE outperforms LS at low SNR, but performs nearly equally as LS when $\rho _{{\text{Con}}}$ is high. Secondly, different from conventional methods, the data-aided channel estimation method makes full use of UL data signal to explore channel information it carries. In addition, in our two-tier network model, decoded UL data streams at the SBSs will be assembled at the central MBS via capacity-rich backhaul links. As such, we can utilize the collected UL data streams as a kind of asymptotically orthogonal pilot sequences, although these sequences have a certain probability of error. As a matter of fact, the data-aided method offers more information of the desired channels and introduces more degrees of freedom in the process of estimation, such as UL data power, UL data length and BER performance. Furthermore, the advantage of our proposed scheme can be clearly shown by an increment in the SNR-like term $\rho _{{\text{DA}}}$ in TABLE \ref{ComTable}, which implies that the data-aided method does help with elevating channel estimation performance in all cases.

Some intuitions can be gained on how to obtain better channel for downlink transmission in decoupled systems using the data-aided channel estimation method. According to the NMSE expression for the data-aided method, we can gain several insights into practical design.
\begin{itemize}
  \item Generally, when we increase training power, training length, UL data length and UL power, NMSE will monotonically decrease and better channel estimation can be achieved.
  \item When we reduce our generalized model to the scenario without UL data BER, it basically corresponds to the case where all UL data are decoded correctly and $\rho _{{\text{DA}}} = \left( \frac{{\tau _{\text{T}} P_{\text{T}} + \tau _{\text{D}} P_{\text{D}}}}{{N_0 }} \right)$. As a result, NMSE is determined by the total energy within training slot and UL data transmission slot, which is different from some previous literatures where power and length distribution within two slots can be optimized by assuming a fixed total energy \cite{hassibi2003much,jindal2010unified,jindal2009value}. However, in our model, NMSE is a constant regardless how to allocate power and length between the two slots. Also, the performance of this zero-BER case can be regarded as the performance upper bound.
  \item If the BER performance is extremely poor in certain circumstances, say close to $0.5$, the benefit from data-aided estimation in  the SNR-like term will disappear and $\rho _k^{{\text{DA}}}  = \frac{{\tau _{\text{T}} P_{\text{T}} }} {{N_0 }}$. The data-aided method will be degraded to the conventional method. Hence, BER is a significant factor to exploit the most of channel information from data sequences. Further, NMSE of a typical UE is not only affected by its own UL data BER, but also by BER of co-channel UEs due to non-orthogonality property of coded data sequences although both sequences are orthogonal asymptotically.
  \item As observed, we can improve the NMSE for all the estimators listed in TABLE I by increasing training power. However, note that the SNR-like term for data-aided estimation would eventually saturate to \vspace{-5pt}
%        \begin{equation}\label{}
%        \frac{{\tau _{\text{D}} \left( {1 - 2{\text{BER}}_{v1} } \right)^2 }}
%        {{\sum\limits_k^K {\beta _k^{\text{M}} \left\{ {1 - \left( {1 - 2{\text{BER}}_{vk} } \right)^2 } \right\}}  + \frac{{N_0 }}
%        {{P_{\text{D}} }}}}
%        \end{equation}
        \begin{equation}\label{NMSE_PdSt}
        \frac{{\tau _{\text{D}} \left( {1 - 2{\text{BER}}_{v1} } \right)^2 }}
        {{\sum\limits_k^K {\beta _k^{\text{M}} \left\{ {1 - \left( {1 - 2{\text{BER}}_{vk} } \right)^2 } \right\}} }}
        \end{equation}
        when ${P_{\text{D}} }$ approaches infinity. Therefore, there is a performance floor when increasing data power, although in practice the maximum transmit power is fixed.
\end{itemize}

\section{Numerical Results}
In this section, we perform simulations to validate the analytical results and demonstrate the potential of data-aided channel estimation for  decoupled UEs. As described in the system model, we consider a single cell consisting of an MBS in the center and $S$ SBSs uniformly placed in the cell. Also, $K$ UEs are uniformly distributed in random within the cell and connect to the BSs according to the modified MARP with decoupling access. As a result, there are a number of decoupled UEs who connect to different BSs in UL/DL. Unless otherwise specified, numerical results provided focus on and are averaged over all decoupled UEs, and the parameters in this section are listed in TABLE \ref{ParaTable}. All points in simulations were obtained via $100$ association patterns and $1000$ independent channel realizations. In this section, PO corresponds to ``pilot only'' and DA represents the ``data-aided'' method.
\begin{table}[!htbp]
\centering
\caption{Simulation Parameters}\label{ParaTable}
\begin{tabular}{ c || l }
  \hline
 \bf{Parameters} & \bf{Values} \\
  \hline
  Radius of macro cell  $R_{\rm{M}}$ & 1000 m \\
  Number of SBSs  $S$ & 30  \\
  Number of UEs  $K$ & 30   \\
  Noise power spectral density  $N_0$ & -174 dBm/Hz   \\
  Attenuation exponent $\alpha$ & 4\\
  Antenna number of MBS $M$ & $256$ \\
  Antenna number of SBS  $N$ & 8   \\
  Transmit power of MBS  $P_{\rm{M}}$ & 46 dBm   \\
  Transmit power of SBS  $P_{\rm{S}}$ & 24 dBm   \\
  Training power of UEs  $P_{\rm{T}}$ & -7$\sim$23 dBm   \\
  Data power of UEs  $P_{\rm{D}}$ & -7$\sim$23 dBm   \\
  Training length  $\tau_{\rm{T}}$ & 30   \\
  Data length  $\tau_{\rm{D}}$ & 128   \\
  \hline
\end{tabular}
\end{table}

Two main performance metrics are used to evaluate the proposed DA method. One is the NMSE defined in Section \uppercase\expandafter{\romannumeral4} and the second performance metric is the average DL rate by assuming standard zero-forcing beamforming (ZFBF) based on the estimated channels at all BSs. The average DL rate is defined as

\begin{equation}\label{}
\mathcal{R} = \mathbb{E}\left[ {\log _2 \left( {1 + {\rm{SINR}}_{{\text{UE}}}^{{\text{DL}}} } \right)} \right],
\end{equation}
where ${\rm{SINR}}_{{\text{UE}}}^{{\text{DL}}}$ is the received DL ${\rm{SINR}}$ of UEs. Note that the maximum UL power of UEs in some figures is extended to 43~dBm to exhibit the whole picture of impacts on metrics, but in practice UL power would not be raised to that high .

In order to verify data-aided scheme can work in other scenarios, NLoS/LoS path loss model is also adopted in simulations. Due to page restrictions, only rate performance is shown and NMSE performance of this model is similar to all-NLoS model. According to 3GPP\cite{3gpp2012} and previous work\cite{ding2016performance}, NLoS/LoS path loss model is defined as
\begin{equation}
\begin{split}
\beta _k^{{\rm{M,NL}}} &= A_{\rm{M}}^{{\rm{NL}}}{\left( {d_k^{\rm{M}}} \right)^{ - \alpha _{\rm{M}}^{{\rm{NL}}}}},\\
\beta _{sk}^{{\rm{S,L}}} &= A_{\rm{S}}^{\rm{L}}{\left( {d_{sk}^{\rm{S}}} \right)^{ - \alpha _{\rm{S}}^{\rm{L}}}}
\end{split}
\end{equation}
where $A_{\rm{M}}^{{\rm{NL}}}$ and $A_{\rm{S}}^{\rm{L}}$ are the path loss of NLoS path between a MBS and a UE and the path loss of LoS path between a SBS and a UE at a reference distance of $1$. ${\alpha _{\rm{M}}^{{\rm{NL}}}}$ and ${\alpha _{\rm{S}}^{\rm{L}}}$ are path loss exponents for their corresponding paths. According to the recommendations of 3GPP, we adopt the parameters as
$\alpha _{\rm{M}}^{{\rm{NL}}} = 3.75,
\alpha _{\rm{S}}^{\rm{L}} = 2.09,
A_{\rm{M}}^{{\rm{NL}}} = {10^{ - 14.54}},
A_{\rm{S}}^{\rm{L}} = {10^{ - 10.38}}.$

%Fig.~\ref{P1_Pt} represents the NMSE performance of conventional methods and as we can see, the simulations match well with our analysis of LS and MMSE estimators for decoupled UEs. Also, MMSE outperforms LS when $\rm{SNR_T}$ is small and becomes almost identical as $\rm{SNR_T}$ grows. Results in this figure further verify that the UL performance for decoupled UEs associated with the SBSs (blue line) is much better (roughly $10$dB) than those connecting to the MBS (red line). This explains not only why those decoupled UEs connect to the SBSs in UL according to the decoupled association policy but also the major reason that channel information quality at the MBS should be improved for better DL performance.
%\begin{figure}
%\centering
%	\includegraphics[width=8cm]{Figures//P1_Pt.pdf}
%    \caption{NMSE versus $\rm{SNR_T}$ with ${P_{\text{D}} }=23$ dBm for LS and MMSE conventional channel estimation methods.  }
%    \label{P1_Pt}
%\end{figure}
\begin{figure}
\begin{minipage}{0.5\textwidth}
\centering
\includegraphics[width=8.5cm]{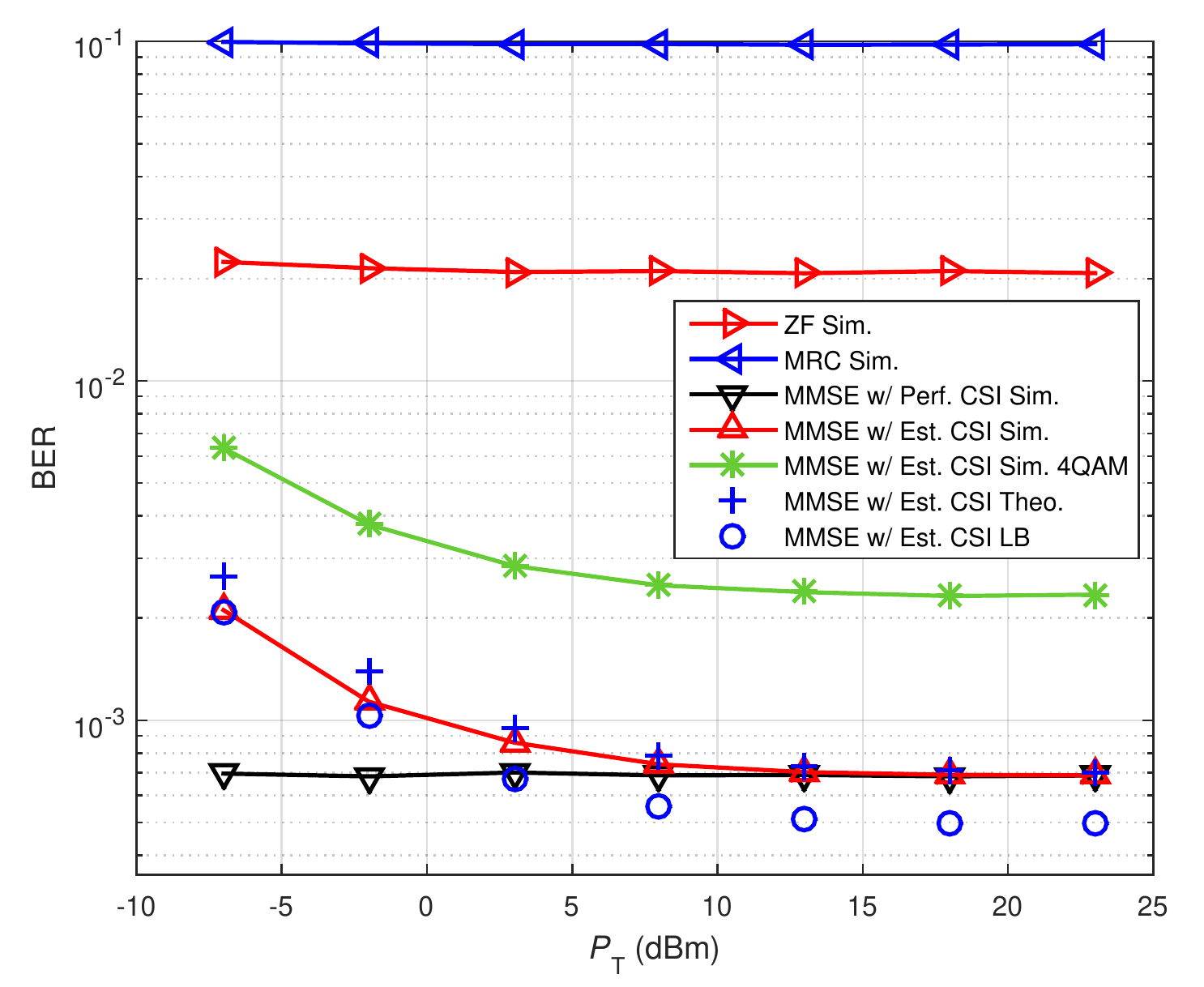}
\caption{Average uplink data BER of decoupled UEs versus ${P_{\text{T}} }$ with ${P_{\text{D}} }=23$ dBm for different data estimators.}
\label{P2_Pt}
\end{minipage}
\begin{minipage}{0.5\textwidth}
\centering
\includegraphics[width=8.5cm]{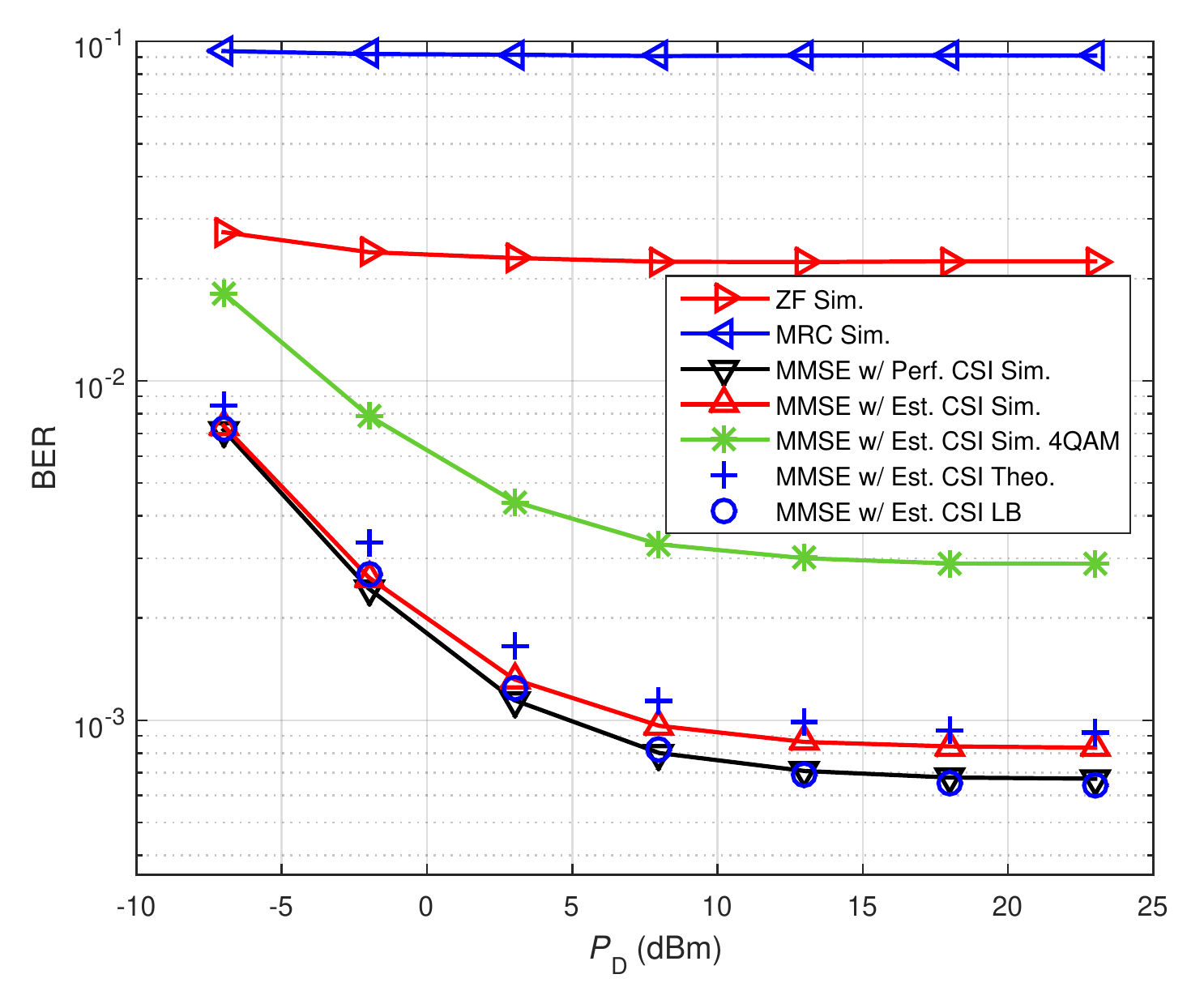}
\caption{Average uplink data BER of decoupled UEs versus ${P_{\text{D}} }$ with ${P_{\text{T}} }=3$ dBm for different data estimators.}
\label{P2_Pd}
\end{minipage}
\end{figure}

In Fig.~\ref{P2_Pt} and Fig.~\ref{P2_Pd}, the average UL data BER performance of decoupled UEs with different detectors versus training power and data power are demonstrated. Numerical results reveal that ZF enjoys better performance due to interference cancelation with CSI at the SBSs, but both detectors are not able to obtain as low BER as MMSE when training or data power grows. All the MMSE estimators with estimated CSI can achieve BER performance better than $10^{-3}$ under severe co-channel interference and the curves for those estimators saturate to the one for the MMSE detector with perfect CSI. The estimated BER with imperfect CSI in (\ref{EstBER}) is plotted with plus signs and is pretty close to the simulated one, while a gap between BER simulations and its lower bound (in circles) is observed in both figures. Hence, the estimated BER is accurate enough to be utilized to represent the real BER performance of the decoded data in the later DA channel estimation. Note that all the curves are for BPSK modulation except the one in asterisk is for the BER performance of 4QAM modulation, which has higher BER than BPSK since the power of UL symbols remains the same in different modulations.

%\begin{figure}
%\centering
%	\includegraphics[width=10cm]{Figures//P2_Pt.pdf}
%    \caption{Average uplink data BER of decoupled UEs versus $\rm{SNR_T}$ with ${P_{\text{D}} }=23$ dBm for different data estimators. }
%    \label{P2_Pt}
%\end{figure}

%\begin{figure}
%\centering
%	\includegraphics[width=10cm]{Figures//P2_Pd.pdf}
%    \caption{Average uplink data BER of decoupled UEs versus $\rm{SNR_D}$ with ${P_{\text{T}} }=3$ dBm for different data estimators. }
%    \label{P2_Pd}
%\end{figure}
\begin{figure}
\begin{minipage}{0.5\textwidth}
\centering
\includegraphics[width=8.5cm]{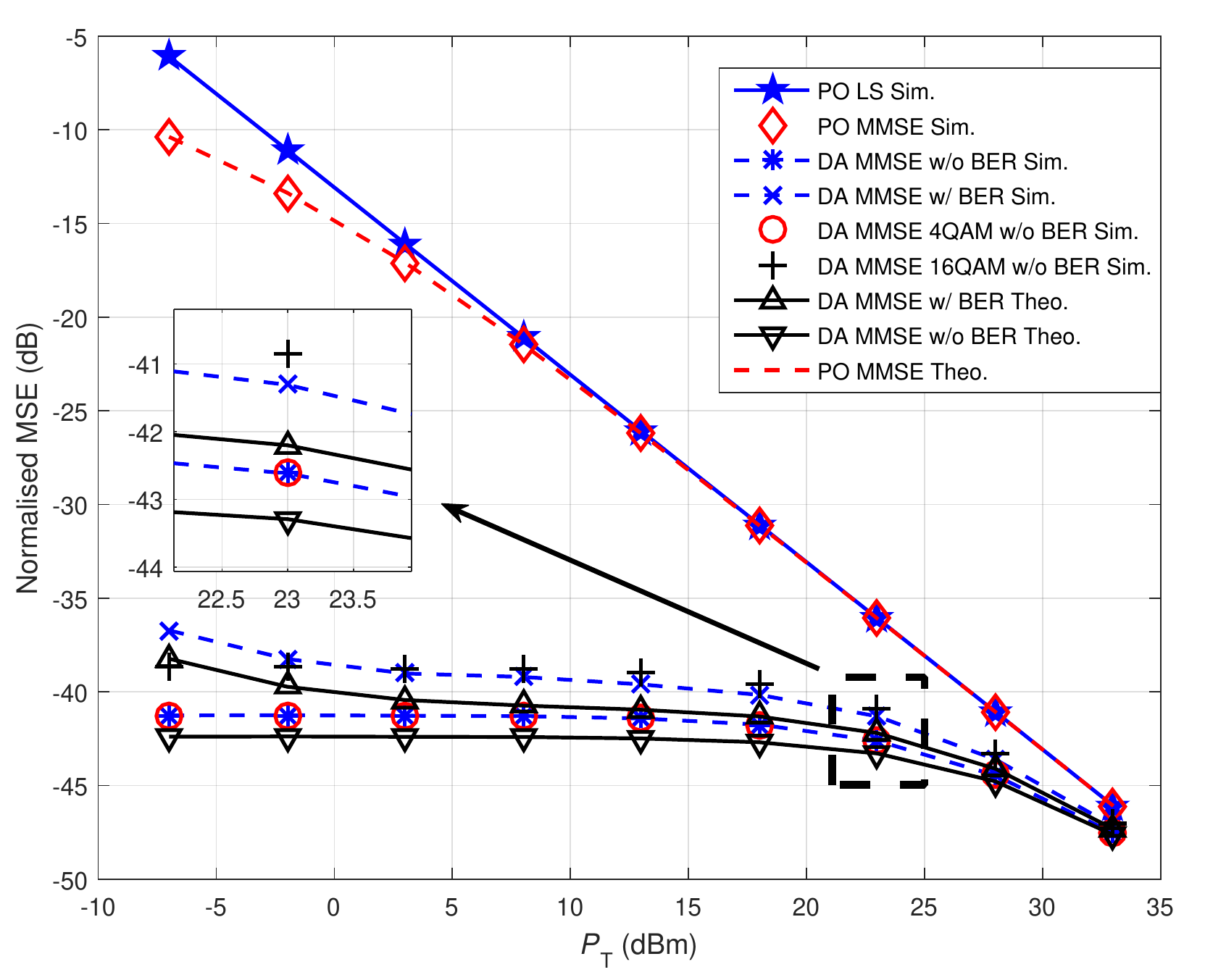}
\caption{Comparison of average NMSE of decoupled UEs for ${P_{\text{T}} }$ with ${P_{\text{D}} }=23$ dBm.}
\label{P3_Pt}
\end{minipage}
\begin{minipage}{0.5\textwidth}
\centering
\includegraphics[width=8.5cm]{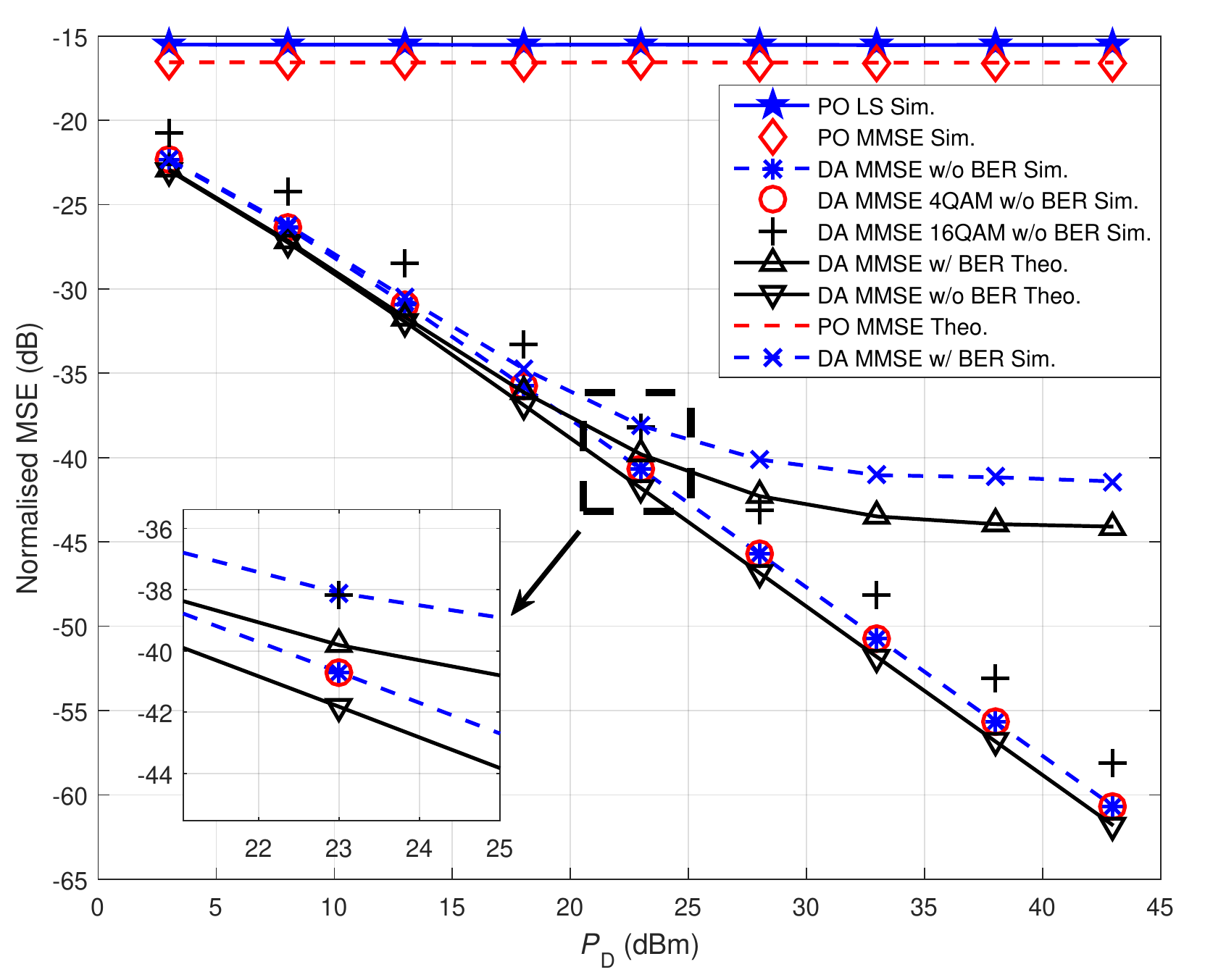}
\caption{Comparison of average NMSE of decoupled UEs for ${P_{\text{D}} }$ with ${P_{\text{T}} }=3$ dBm.}
\label{P3_Pd}
\end{minipage}
\end{figure}

Then, we verify the NMSE performance of the DA method and compare it with the PO LS and MMSE. Fig.~\ref{P3_Pt} depicts the relation between training power and NMSE of decoupled UEs. It is observed that the DA channel estimation is always better than LS and MMSE estimators, in particular, when training power is low, the DA method outperforms the conventional ones over 30dB. The curves for the DA method tend to approach to and eventually saturate to the one for the PO MMSE and LS estimators as training power increases, which can be explain with the NMSE expression for DA in TABLE \ref{ComTable} that when training power grows higher, the first term with training power dominates the SNR-like term which is highly related to the NMSE performance and the benefit from DA method is not as remarkable as in the low training power scenarios. Moreover, the squares represent the NMSE by treating a decoded sequence with errors as if there were no error in it. Its performance is almost the same as the one considering the BER effect at low training power regime and become worse as training power grows. Interestingly, in the error-free scenario, we find the NMSE performance of 4QAM is exactly the same as that of BPSK and both two modulations performs a little better than 16QAM, which provides a solid proof for the insight concluded in Section \uppercase\expandafter{\romannumeral4}-C that NMSE of the DA method is determined by the total energy of training and data slots since all symbols of 4QAM and BPSK have the same power ${P_{\text{T}} }$ while symbols of 16QAM have several power levels with the maximum power of ${P_{\text{T}} }$.
%\begin{figure}
%\centering
%	\includegraphics[width=10cm]{Figures//P3_Pt.pdf}
%    \caption{Comparison of average NMSE of decoupled UEs for $\rm{SNR_T}$ with ${P_{\text{D}} }=23$ dBm.  }
%    \label{P3_Pt}
%\end{figure}

The NMSE versus data power is shown in Fig.~\ref{P3_Pd}. The gap between PO and the DA methods gets larger by increasing data power for the reason that the NMSE performances of PO methods has no relation with data power while for the DA method, higher data power will not only improve BER performance but also directly increase the SNR-like term. Furthermore, the NMSE of the DA method without BER can be improved almost log-linearly as data power grows while the case with BER tends to decrease slower and reaches a performance limit as analyzed in Section \uppercase\expandafter{\romannumeral4}-C. Similar results of 4QAM and 16QAM can be observed in this figure.
%\begin{figure}
%\centering
%	\includegraphics[width=10cm]{Figures//P3_Pd.pdf}
%    \caption{Comparison of average NMSE of decoupled UEs for $\rm{SNR_D}$ with ${P_{\text{T}} }=3$ dBm.   }
%    \label{P3_Pd}
%\end{figure}
\begin{figure}
\begin{minipage}{0.5\textwidth}
\centering
\includegraphics[width=8.5cm]{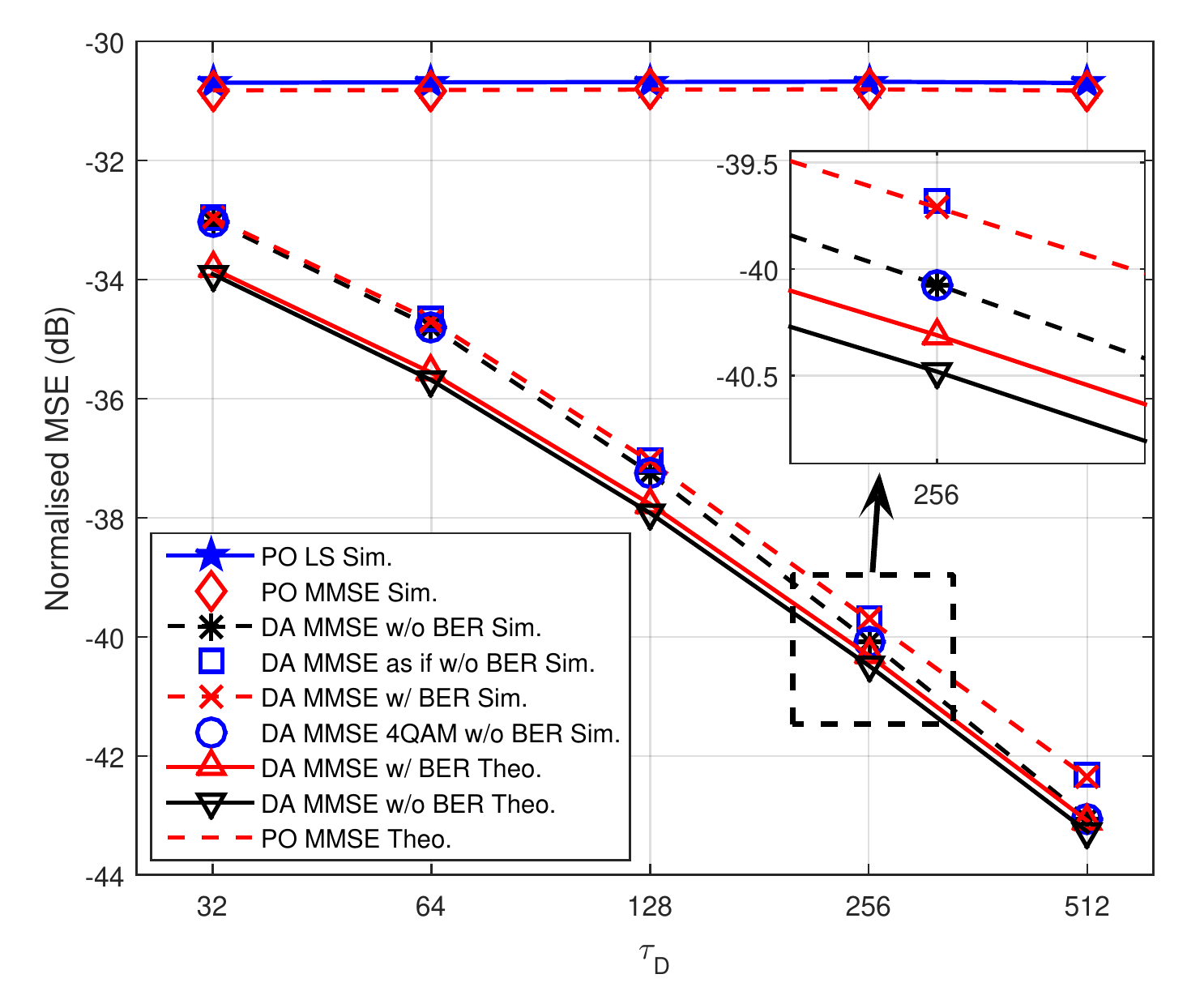}
\caption{Comparison of average NMSE of decoupled UEs for $\tau_{\rm{D}}=64,128,256,512$ with ${P_{\text{T}} }=13$ dBm and ${P_{\text{D}} }=13$ dBm.}
\label{P3_Td}
\end{minipage}
\begin{minipage}{0.5\textwidth}
\centering
\includegraphics[width=8.5cm]{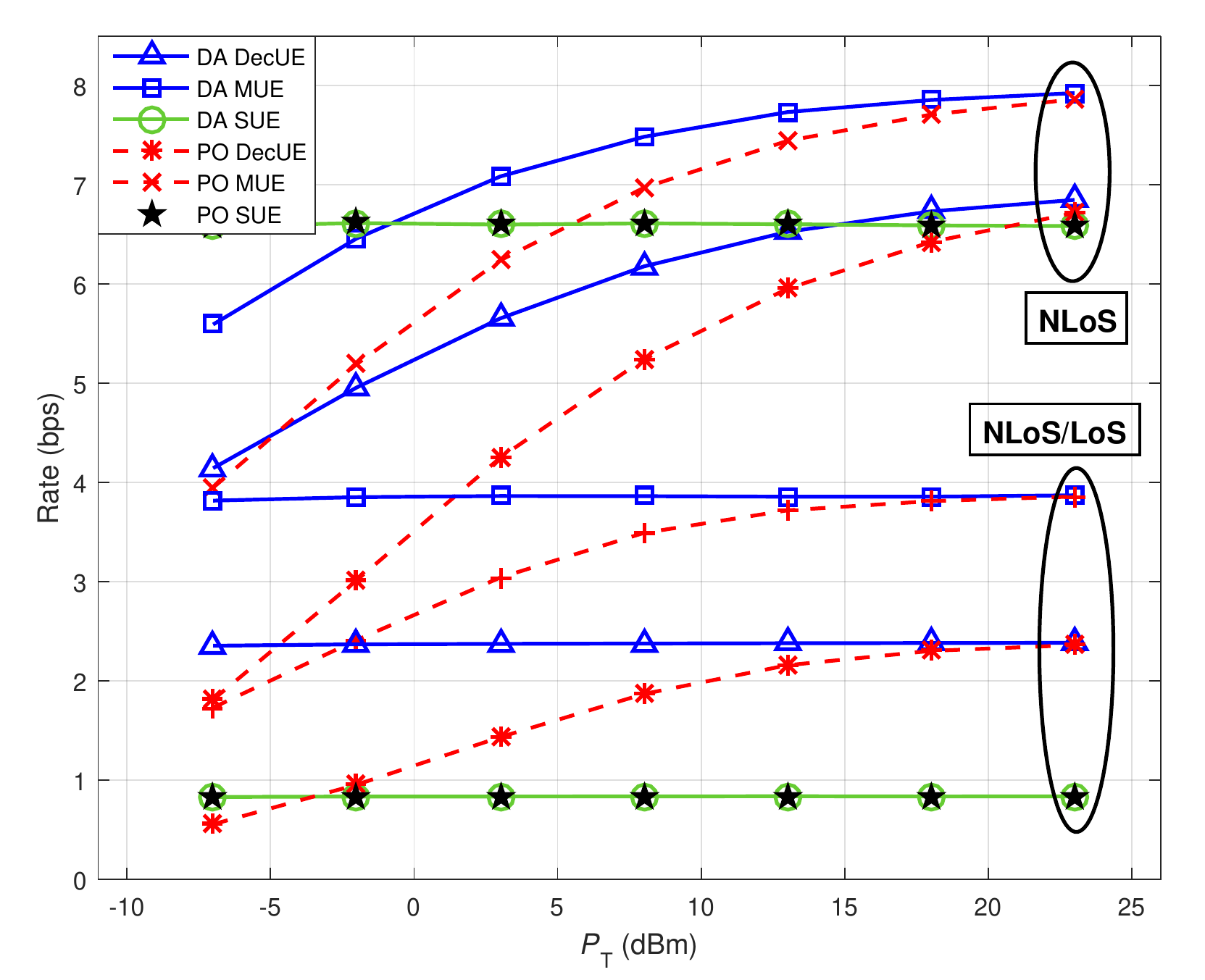}
\caption{Average DL rate versus ${P_{\text{T}} }$ with ${P_{\text{D}} }=23$ dBm for the  comparison between DA and PO.}
\label{P4_Pt}
\end{minipage}
\end{figure}

The relation between NMSE and data lengths is also investigated. In Fig.~\ref{P3_Td}, it is observed that when the UL data length gets longer, NMSE for the PO MMSE remains constant but the NMSE for the DA method enjoys nearly log-linear improvement. Recall that we make the approximation that ${\mathbf{\hat X\hat X}}^{\text{H}}  \to \tau _{\text{D}} P_{\text{D}} {\mathbf{I}}_k$ when $\tau _{\text{D}}$ is large. Fig.~\ref{P3_Td} indicates that for the DA method without BER, the gap between simulations and the approximated result is getting smaller quickly when $\tau _{\text{D}}$ becomes larger, while for the case with BER, the approximation gap reduces slowly. In general, as the result of BER effect, the DA methods without BER performs better than the method with BER and the performance gap between two methods gets larger as the length of data grows.

\begin{figure}
\begin{minipage}{0.5\textwidth}
\centering
\includegraphics[width=8.5cm]{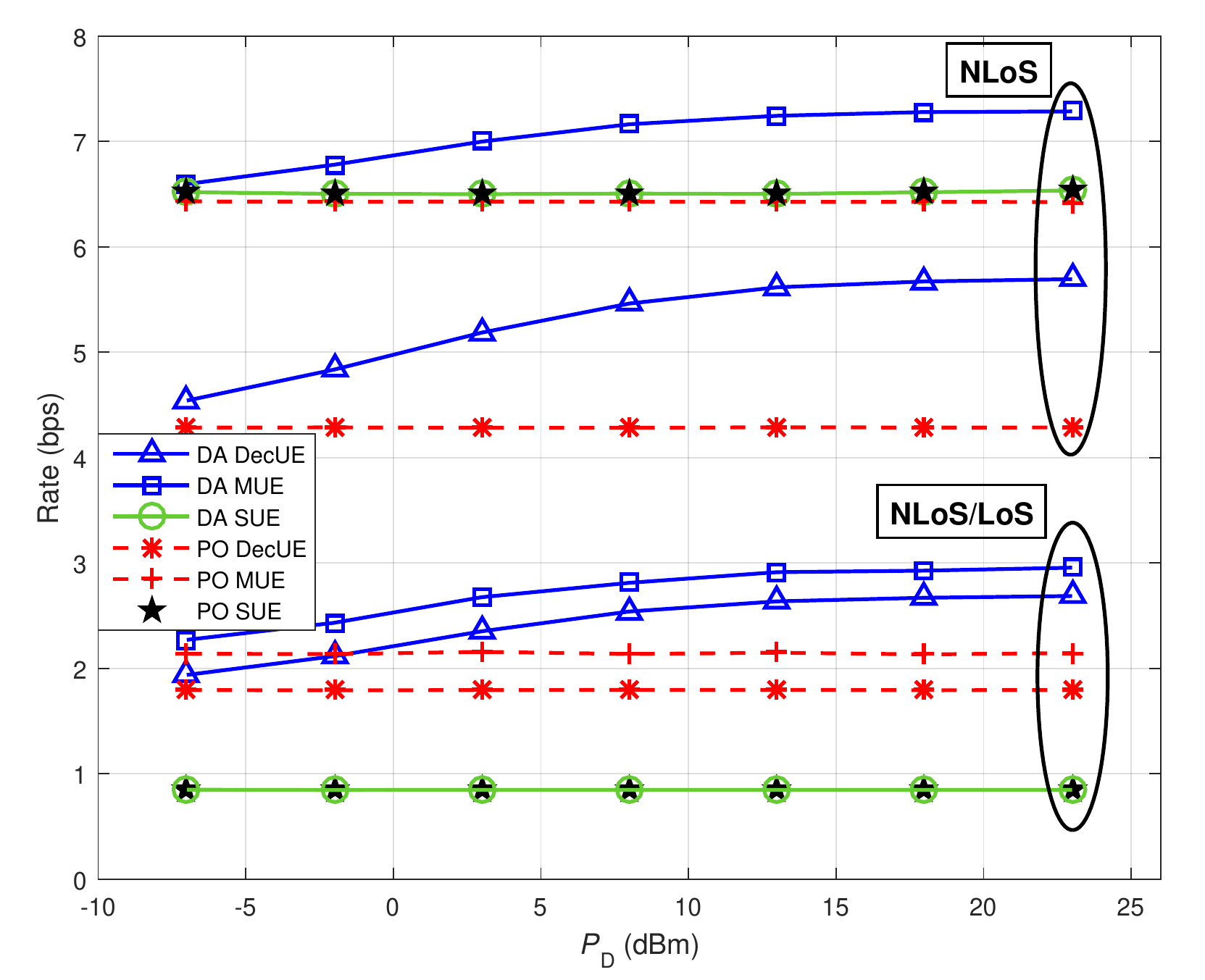}
\caption{Average DL rate versus ${P_{\text{D}} }$ with ${P_{\text{T}} }=3$ dBm for the comparison between DA and PO.}
\label{P4_Pd}
\end{minipage}
\begin{minipage}{0.5\textwidth}
\centering
\includegraphics[width=8.5cm]{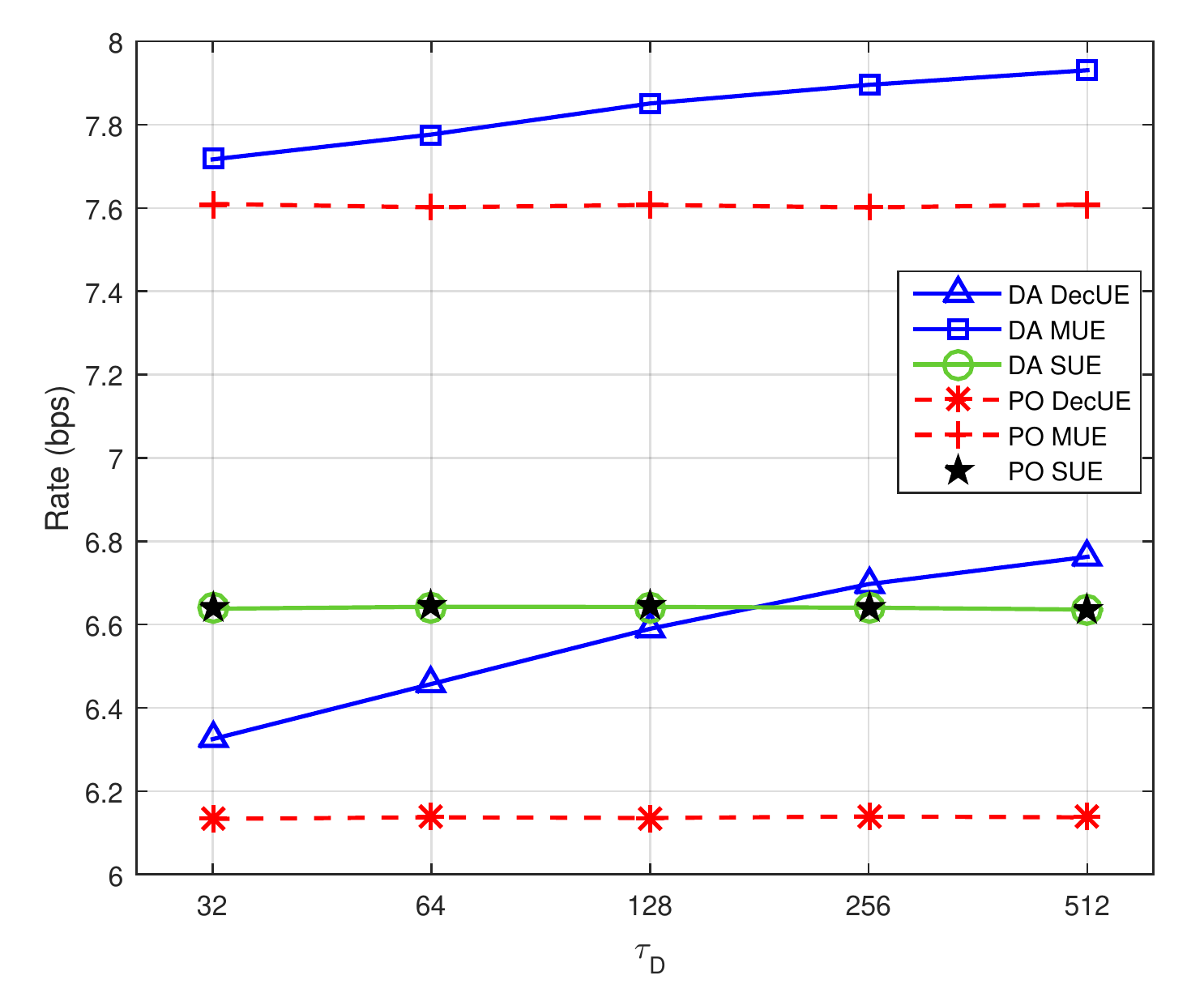}
\caption{Average DL rate versus ${\tau_{\text{T}} }$ with ${P_{\text{T}} }=3$ dBm and ${P_{\text{D}} }=23$ dBm for the comparison between DA and PO.}
\label{P4_Td}
\end{minipage}
\end{figure}

Next, we examine the impacts of parameters on the final rate metric and compare the average DL rate performance between three types of UEs, which are decoupled UEs, SUEs (served by SBSs in UL/DL) and MUEs (served by the MBS in UL/DL), also between two different path loss models. From Fig.~\ref{P4_Pt} and Fig.~\ref{P4_Pd}, we observe that DA method can improve DL rate of both MUEs and decoupled UEs by using two different path loss models within the whole range of training and data power, especially in high data power and low training power regimes. By utilizing DA method, the MBS has more accurate channels of decoupled UEs and the accuracy of DL beamforming is also improved, thus, the inter-UE interference at both MUEs and decoupled UEs can be suppressed effectively and higher rate can be obtained. The average rate for SUEs remains almost the same because the UL channel estimation at SBSs is so good that the improvement in NMSE by increasing training power would have little benefit to the DL rate, while it is easy to figure out that changing UL data power has no effect on DL rate of SUEs. Therefore, the DA methods can not only improve the average DL rate of the whole cell but also reduce rate gaps between different types of UEs, which promises better fairness in the cell. Moreover, although DL rates for three types of UEs have dropped by using NLoS/LoS model, the data-aided scheme could still provide remarkable promotion in rate performances of MUEs and decoupled UEs.

Fig.~\ref{P4_Td} shows that the DL rate gain achieved by the DA method grows with the increase of UL data length. Similarly, the DL rates of both MUEs and decoupled UEs are improved by DA method and it is observed that decoupled UEs enjoys larger rate growth than MUEs.

\section{Conclusion}
In this paper, we proposed a data-aided channel estimation method for decoupled UEs in HetNets. In this method, the decoded UL data and known training sequence are jointly utilized to better estimate the DL channels at the MBS. The ergodic BER of UL data has been analyzed to model the decoded data with errors and the approximated NMSE for the data-aided MMSE estimator was derived and compared with the conventional estimators. It has been proved by both theoretic results and simulations that the data-aided method can greatly improve the accuracy of estimated channels of decoupled UEs by introducing more degrees of freedom. Although the BER may affect the effectiveness of this method, the BER could be controlled at a very low level by applying more advanced data detection techniques in the future. DL rate performance is also investigated numerically to show the true benefit to this final metric and the results of 4QAM/16QAM and NLoS/NLoS verify the proposed method can be applied in multiple scenarios with different modulations.

\appendices
\section{Proof of Lemma 1}
\begin{IEEEproof}
From the expression of ${{\rm{SIN}}{{\rm{R}}_{vk}}}$ in (\ref{SINRSplit}), we have
\begin{equation}\label{}
\begin{split}
\mathbb{E}\left[ {{\text{SINR}}_{vk} } \right]
&= \mathbb{E}\left[ {\rho _v \sum\limits_{i = 1}^N {\frac{{\left\| {\varphi_i } \right\|^2 }}
{{1 + \rho _v d_i^2 }}} } \right]\\
%= \rho _v\mathbb{E}\left[ {\left\| {\varphi_i } \right\|^2 } \right]\mathbb{E}\left[ { \sum\limits_{i = 1}^N {\lambda _i} } \right]
&= N\rho _v \hat \beta _{vk}^{\text{S}} \mathbb{E}\left[ {\left( {\frac{1}
{N}\sum\limits_{i = 1}^N {\lambda _i } } \right)} \right]\\
&= N\rho _v \hat \beta _{vk}^{\text{S}} \mathbb{E}\left[ {\frac{{{\text{Tr}}\left( {\mathbf{\Lambda }} \right)}}
{N}} \right]
\approx N\rho _v \hat \beta _{vk}^{\text{S}} \mu,
\end{split}
\end{equation}
where the above approximation uses the definition in (\ref{DefESINR}).

Similarly, we obtain the second moment by
\begin{align}\label{}
{\text{Var}}\left[ {{\text{SINR}}_{vk} } \right]
&= N^2 \rho _v^2 {\text{Var}}\left[ {\frac{1}
{N}\sum\limits_{i = 1}^N {\lambda _i \left\| {\phi _i } \right\|^2 } } \right] \notag\\
&= N^2 \rho _v^2 {\text{Var}}\left[ {\frac{1}
{N}\sum\limits_{i = 1}^N {\lambda _i \mathbb{E}\left[ {\left\| {\phi _i } \right\|^2 |{\mathbf{\hat G}}_{v\left( { - k} \right)} } \right]} } \right] + N^2 \rho _v^2 \mathbb{E}\left[ {\frac{1}
{N}\sum\limits_{i = 1}^N {\lambda _i^2 } {\text{Var}}\left[ {\left\| {\phi _i } \right\|^2 |{\mathbf{\hat G}}_{v\left( { - k} \right)} } \right]} \right] \notag\\
&\overset{(a)}{=}  N^2 \rho _v^2 \left( {\hat \beta _{vk}^{\text{S}} } \right)^2 {\text{Var}}\left[ {\frac{{{\text{Tr}}\left( {\mathbf{\Lambda }} \right)}}
{N}} \right] + N^2 \rho _v^2 \left( {\hat \beta _{vk}^{\text{S}} } \right)^2 \mathbb{E}\left[ {\frac{{{\text{Tr}}\left( {{\mathbf{\Lambda }}^2 } \right)}}
{N}} \right] \notag\\
&\approx N^2 \rho _v^2 \left( {\hat \beta _{vk}^{\text{S}} } \right)^2 \sigma ^2,
\end{align}
where the first term of $(a)$ can be proved to converge to $0$ by the results from \cite{li2006distribution}.
\end{IEEEproof}

\section{Proof of Proposition 2}
\begin{IEEEproof}
First, denote the mean-square-error (MSE) as $\mathcal{J}\left( {{\mathbf{C}}_k } \right) = \mathbb{E}\left[ {\left\| {{\mathbf{\hat h}}_k^{{\text{MMSE}}}  - {\mathbf{h}}_k } \right\|^2 } \right]$. Then by applying the MMSE rule directly, the problem is formulated as
\begin{equation}\label{}
\mathop {\min }\limits_{{\mathbf{C}}_k } \mathcal{J}\left( {{\mathbf{C}}_k } \right) = \\ \mathbb{E}\left[ {\left( {\left( {{\mathbf{H}}\left( {{\mathbf{\hat W}} \circ {\mathbf{E}}} \right) + {\mathbf{Z}}} \right){\mathbf{C}}_k  - {\mathbf{h}}_k } \right)^{\text{H}} \left( {\left( {{\mathbf{H}}\left( {{\mathbf{\hat W}} \circ {\mathbf{E}}} \right) + {\mathbf{Z}}} \right){\mathbf{C}}_k  - {\mathbf{h}}_k } \right)} \right],
\end{equation}
and we take the derivative of ${\mathcal J}$ in terms of ${{\bf{C}}_{k}}$ and let the derivative be ${\bf{0}}$ to find the optimal solution
\begin{equation}\label{MMSE_DA_Exp}
\left( {\mathbb{E}\left[ {\left( {{\mathbf{\hat W}} \circ {\mathbf{E}}} \right)^{\text{H}} {\mathbf{H}}^{\text{H}} {\mathbf{H}}\left( {{\mathbf{\hat W}} \circ {\mathbf{E}}} \right) + {\mathbf{Z}}^{\text{H}} {\mathbf{Z}}} \right]} \right){\mathbf{C}}_k  = \mathbb{E}\left[ {\left( {{\mathbf{\hat W}} \circ {\mathbf{E}}} \right)^{\text{H}} {\mathbf{H}}^{\text{H}} {\mathbf{h}}_k } \right].
\end{equation}

Then according to the statistical properties of ${\mathbf{H}}$, ${\mathbf{E}}$ and the independency between two random matrices, we can obtain
\begin{equation}\label{}
{\mathbf{R}}_{\mathbf{H}}
 \triangleq \frac{1}
{M}\mathbb{E}\left[ {{\mathbf{H}}^{\text{H}} {\mathbf{H}}} \right] = {\text{diag}}\left( {\beta _1^{\text{M}} , \dots ,\beta _k^{\text{M}} , \dots ,\beta _K^{\text{M}} } \right),
\end{equation}
and
\begin{equation}\label{Exp_DAMMSE}
\left( {\mathbb{E}\left[ {\left( {{\mathbf{\hat W}} \circ {\mathbf{E}}} \right)^{\text{H}} {\mathbf{R}}_{\mathbf{H}} \left( {{\mathbf{\hat W}} \circ {\mathbf{E}}} \right)} \right] + N_0 {\mathbf{I}}} \right){\mathbf{C}}_k  = 
\left( {{\mathbf{\hat w}}_k^{\text{H}}  \circ \mathbb{E}\left[ {{\mathbf{e}}_k^{\text{H}} } \right]} \right)\beta _k^{\text{M}}.
\end{equation}

The above expectation can be defined by ${\mathbf{P}}$ and decomposed into four block matrices
\begin{equation}\label{}
{\mathbf{P}}  \triangleq \mathbb{E}\left[ {\left( {\left[ {{\mathbf{S}} \circ {\mathbf{E}}_1 ,{\mathbf{\hat X}} \circ {\mathbf{E}}_2 } \right]} \right)^{\text{H}} {\mathbf{R}}_{\mathbf{H}} \left( {\left[ {{\mathbf{S}} \circ {\mathbf{E}}_1 ,{\mathbf{\hat X}} \circ {\mathbf{E}}_2 } \right]} \right)} \right] 
= \left( {\begin{array}{*{20}c}
   {{\mathbf{P}}_{11} } & {{\mathbf{P}}_{12} }  \\
   {{\mathbf{P}}_{21} } & {{\mathbf{P}}_{22} }   \\
 \end{array} } \right),
\end{equation}
with
${\mathbf{P}}_{11}  = {\mathbf{S}}^{\text{H}} {\mathbf{R}}_{\mathbf{H}} {\mathbf{S}}$,
${\mathbf{P}}_{12}  = {\mathbf{S}}^{\text{H}} {\mathbf{R}}_{\mathbf{H}} \left( {{\mathbf{\hat X}} \circ \mathbb{E}\left[ {{\mathbf{E}}_2 } \right]} \right)$,
${\mathbf{P}}_{21}  = \left( {{\mathbf{\hat X}} \circ \mathbb{E}\left[ {{\mathbf{E}}_2 } \right]} \right)^{\text{H}} {\mathbf{R}}_{\mathbf{H}} {\mathbf{S}}$,\\
${\mathbf{P}}_{22}  = \mathbb{E}\left[ {\left( {{\mathbf{\hat X}} \circ {\mathbf{E}}_2 } \right)^{\text{H}} {\mathbf{R}}_{\mathbf{H}} \left( {{\mathbf{\hat X}} \circ {\mathbf{E}}_2 } \right)} \right]$.

The results of ${\mathbf{P}}_{11}, {\mathbf{P}}_{12}$ and ${\mathbf{P}}_{21}$ are straightforward. Hence, we focus on the expectation of ${\mathbf{P}}_{22}$. Two cases are discussed separately by using the definition of matrix multiplication and independency, as
\begin{align}\label{}
\left[ {{\mathbf{P}}_{22} } \right]_{ij,i \ne j}
&= \sum\limits_k^K {\sum\limits_l^K {\left[ {{\mathbf{R}}_{\mathbf{H}} } \right]_{kl} \left( {\hat x_{ki} } \right)^\prime  } } \hat x_{lj} \mathbb{E}\left[ {\left( {\left[ {{\mathbf{E}}_2 } \right]_{ki} } \right)^\prime  \left[ {{\mathbf{E}}_2 } \right]_{lj} } \right] \notag\\
%&= \sum\limits_k^K {\beta _k^{\text{M}} \left( {\hat x_{ki} } \right)^\prime  \hat x_{kj} \mathbb{E}\left[ {\left( {\left[ {{\mathbf{E}}_2 } \right]_{ki} } \right)^\prime  } \right]} \mathbb{E}\left[ {\left[ {{\mathbf{E}}_2 } \right]_{kj} } \right] \notag\\
&= \sum\limits_k^K {\beta _k^{\text{M}} \left( {\hat x_{ki} } \right)^\prime  \hat x_{kj} } \left( {1 - 2{\text{BER}}_{vk} } \right)^2\notag\\
\left[ {{\mathbf{P}}_{22} } \right]_{ij,i = j}
&= \sum\limits_k^K {\beta _k^{\text{M}} \left( {\hat x_{ki} } \right)^\prime  \hat x_{ki} \mathbb{E}\left[ {\left\| {\left[ {{\mathbf{E}}_2 } \right]_{ki} } \right\|^2 } \right]}\notag
%&= \sum\limits_k^K {\beta _k^{\text{M}} \left( {\hat x_{ki} } \right)^\prime  \hat x_{ki}}  \notag\\
= P_{\text{D}} \sum\limits_k^K {\beta _k^{\text{M}} }.
\end{align}

Therefore, $\left[ {{\mathbf{P}}_{22} } \right]_{i \ne j}$ can be generally written as \begin{small}$
\left[ {\left( {\left[ {{\mathbf{S}},{\mathbf{\hat X}} \circ \mathbb{E}\left[ {{\mathbf{E}}_2 } \right]} \right]} \right)^{\text{H}} {\mathbf{R}}_{\mathbf{H}} \left( {\left[ {{\mathbf{S}},{\mathbf{\hat X}} \circ \mathbb{E}\left[ {{\mathbf{E}}_2 } \right]} \right]} \right)} \right]_{i \ne j}$\end{small}, but diagonal elements of ${{\mathbf{P}}_{22} }$ are $
P_{\text{D}} \sum\limits_k^K {\beta _k^{\text{M}} }$. Hence, for brevity and ease of calculation, we can rewrite $
{{\mathbf{P}}_{22} }$ as
\begin{equation}\label{}
\left( {{\mathbf{\hat X}} \circ \mathbb{E}\left[ {{\mathbf{E}}_2 } \right]} \right)^{\text{H}} {\mathbf{R}}_{\mathbf{H}} \left( {{\mathbf{\hat X}} \circ \mathbb{E}\left[ {{\mathbf{E}}_2 } \right]} \right) + \Delta {\mathbf{P}}_{\mathbf{X}},
\end{equation}
where \begin{small}$\Delta {\mathbf{P}}_{\mathbf{X}}=\Delta S_{\mathbf{X}} {\mathbf{I}}_{\tau _{\text{D}} }$,$\Delta S_{\mathbf{X}}=P_{\text{D}} \sum\limits_k^K {\beta _k^{\text{M}} \left\{ {1 - \left( {1 - 2{\text{BER}}_{vk} } \right)^2 } \right\}}$\end{small}.
As a result, the block matrices can be reunited as the sum of two parts denoted as ${\mathbf{P}} \triangleq {\mathbf{\hat P}} + \Delta {\mathbf{P}}$, with
\begin{equation}\label{}
{\mathbf{\hat P}} = \left( {\left[ {{\mathbf{S}},{\mathbf{\hat X}} \circ \mathbb{E}\left[ {{\mathbf{E}}_2 } \right]} \right]} \right)^{\text{H}} {\mathbf{R}}_{\mathbf{H}} \left( {\left[ {{\mathbf{S}},{\mathbf{\hat X}} \circ \mathbb{E}\left[ {{\mathbf{E}}_2 } \right]} \right]} \right)
\end{equation}
and
\begin{equation}
\Delta {\mathbf{P}} = \left( {\begin{array}{*{20}c}
   {\mathbf{0}} & {\mathbf{0}}  \\
   {\mathbf{0}} & {\Delta {\mathbf{P}}_{\mathbf{X}} }  \\
 \end{array} } \right).
\end{equation}

By plugging these results into (\ref{Exp_DAMMSE}), the optimal combination matrix for the data-aided channel estimation method with the MMSE estimator is obtained. After the combination matrix is obtained, we can continue to evaluate NMSE of the data-aided method. From the definition in (\ref{NMSE_Criterio}), we can first calculate numeric term inside logarithm as\vspace{-5pt}
\begin{equation}\label{NMSEMMSEAppendix}
\begin{split}
&\mathbb{E}\left[ {\left\| {{\mathbf{\hat h}}_k^{{\text{MMSE}}}  - {\mathbf{h}}_k } \right\|^2 } \right] \\
= &\mathbb{E}\left[ {\left( {\left( {{\mathbf{H}}\left( {{\mathbf{\hat W}} \circ {\mathbf{E}}} \right) + {\mathbf{Z}}} \right){\mathbf{C}}_k  - {\mathbf{h}}_k } \right)^{\text{H}} \left( {\left( {{\mathbf{H}}\left( {{\mathbf{\hat W}} \circ {\mathbf{E}}} \right) + {\mathbf{Z}}} \right){\mathbf{C}}_k  - {\mathbf{h}}_k } \right)} \right] \\
 = &M\beta _k^{\text{M}}  - M\left( {\beta _k^{\text{M}} } \right)^2 \mathbb{E}\left[ {\left( {{\mathbf{\hat w}}_k  \circ \mathbb{E}\left[ {{\mathbf{e}}_k } \right]} \right)\left( {{\mathbf{P}} + N_0 {\mathbf{I}}} \right)^{ - 1} \left( {{\mathbf{\hat w}}_k^{\text{H}}  \circ \mathbb{E}\left[ {{\mathbf{e}}_k^{\text{H}} } \right]} \right)} \right]
\end{split}
\end{equation}
\vspace{-5pt}

Now, we denote
\begin{equation}\label{}
{\mathbf{P}}_{{\text{Aux}}}   \triangleq  \Delta {\mathbf{P}} + N_0 {\mathbf{I}} = \left( {\begin{array}{*{20}c}
   {N_0 {\mathbf{I}}_{\tau _{\text{T}} } } & {\mathbf{0}}  \\
   {\mathbf{0}} & { \left(\Delta S_{\mathbf{X}}  + N_0\right ) {\mathbf{I}}_{\tau _{\text{D}} } }  \\
 \end{array} } \right)
\end{equation}
and
\begin{equation}\label{}
{\mathbf{\bar W}} \triangleq {\mathbf{\hat W}} \circ \mathbb{E}\left[ {\mathbf{E}} \right].
\end{equation}

Thus, we can derive the expectation part in the last equation of (\ref{NMSEMMSEAppendix}) as
\begin{equation}\label{Exp_MSEMMSE}
\begin{split}
&\mathbb{E}\left[ {\left( {{\mathbf{\hat w}}_k  \circ \mathbb{E}\left[ {{\mathbf{e}}_k } \right]} \right)\left( {{\mathbf{P}} + N_0 {\mathbf{I}}} \right)^{ - 1} \left( {{\mathbf{\hat w}}_k^{\text{H}}  \circ \mathbb{E}\left[ {{\mathbf{e}}_k^{\text{H}} } \right]} \right)} \right] \\
%= &\mathbb{E}\left[ {\left[ {\left( {{\mathbf{\hat W}} \circ \mathbb{E}\left[ {\text{E}} \right]} \right)\left( {\hat{\mathbf{P}} + \Delta {\mathbf{P}} + N_0 {\mathbf{I}}} \right)^{ - 1} \left( {{\mathbf{\hat W}}^{\text{H}}  \circ \mathbb{E}\left[ {{\text{E}}^{\text{H}} } \right]} \right)} \right]_{kk} } \right] \\
= &\mathbb{E}\left[ {\left[ {{\mathbf{\bar W}}\left( {{\mathbf{\bar W}}^{\text{H}} {\mathbf{R}}_{\mathbf{H}} {\mathbf{\bar W}} + {\mathbf{P}}_{{\text{Aux}}} } \right)^{ - 1} {\mathbf{\bar W}}^{\text{H}} } \right]_{kk} } \right] \\
\overset{(a)}= &\mathbb{E}
\bigg[ \Big[ {\mathbf{\bar WP}}_{{\text{Aux}}}^{ - 1} {\mathbf{\bar W}}^{\text{H}}
- {\mathbf{\bar WP}}_{{\text{Aux}}}^{ - 1} {\mathbf{\bar W}}^{\text{H}} \left( {{\mathbf{R}}_{\mathbf{H}}^{ - 1}  + {\mathbf{\bar WP}}_{{\text{Aux}}}^{ - 1} {\mathbf{\bar W}}^{\text{H}} } \right)^{ - 1}{\mathbf{\bar WP}}_{{\text{Aux}}}^{ - 1} {\mathbf{\bar W}}^{\text{H}}  \Big]_{kk}  \bigg]
\\
\overset{(b)}= &\left[ {{\mathbf{R}}_{\mathbf{H}}^{ - 1}  - {\mathbf{R}}_{\mathbf{H}}^{ - 1} \mathbb{E}\left[ {\left( {{\mathbf{\bar WP}}_{{\text{Aux}}}^{ - 1} {\mathbf{\bar W}}^{\text{H}}  + {\mathbf{R}}_{\mathbf{H}}^{ - 1} } \right)^{ - 1} } \right]{\mathbf{R}}_{\mathbf{H}}^{ - 1} } \right]_{kk},
\end{split}
\end{equation}
where (a) follows the Woodbury matrix inversion identity while (b) uses the matrix identity
\begin{equation}\label{}
{\mathbf{A}} - {\mathbf{A}}\left( {{\mathbf{A}} + {\mathbf{B}}} \right)^{ - 1} {\mathbf{A}} = {\mathbf{B}} - {\mathbf{B}}\left( {{\mathbf{A}} + {\mathbf{B}}} \right)^{ - 1} {\mathbf{B}}.
\end{equation}

With the definitions of\vspace{-5pt}
\begin{equation}\label{}
{\mathbf{\tilde E}}_2 \triangleq{\text{ diag}}\left( {\left( {1 - 2{\text{BER}}_{v1} } \right), \dots ,\left( {1 - 2{\text{BER}}_{vK} } \right)} \right)
\end{equation}
and
\begin{equation}\label{}
{\mathbf{\Lambda }}_{\mathbf{X}}
\triangleq \frac{1}
{{\Delta S_{\mathbf{X}}  + N_0 }}{\mathbf{\hat X\hat X}}^{\text{H}},
\end{equation}
the expectation part in (\ref{Exp_MSEMMSE}) can be simplified as
\begin{align}\label{}
&\mathbb{E}\left[ {\left( {{\mathbf{\bar WP}}_{{\text{Aux}}}^{ - 1} {\mathbf{\bar W}}^{\text{H}}  + {\mathbf{R}}_{\mathbf{H}}^{ - 1} } \right)^{ - 1} } \right] \notag\\
= &\mathbb{E}\left[ {\left( {\left[ {{\mathbf{S}},{\mathbf{\hat X}} \circ \mathbb{E}\left[ {{\mathbf{E}}_2 } \right]} \right]\Delta {\mathbf{P}}_{{\text{Aux}}}^{ - 1} \left[ {\begin{array}{*{20}c}
   {{\mathbf{S}}^{\text{H}} }  \\
   {\left( {{\mathbf{\hat X}} \circ \mathbb{E}\left[ {{\mathbf{E}}_2 } \right]} \right)^{\text{H}} }  \\
\end{array} } \right] + {\mathbf{R}}_{\mathbf{H}}^{ - 1} } \right)^{ - 1} } \right] \notag\\
%= &\mathbb{E}\left[ {\left( {\frac{{\tau _{\text{T}} P_{\text{T}} }}
%{{N_0 }}{\mathbf{I}}_k  + \left( {{\mathbf{\hat X}}{\mathbf{\tilde E}}_2 } \right)\left( {\begin{array}{*{20}c}
%   {\frac{1}
%{{N_0 }}{\mathbf{I}}_{\tau _{\text{T}} } } & {\mathbf{0}}  \\
%   {\mathbf{0}} & {\Delta {\mathbf{P}}_{\mathbf{X}}^{ - 1} }  \\
% \end{array} } \right)\left( {{\mathbf{\hat X}}{\mathbf{\tilde E}}_2 } \right)^{\text{H}}  + {\mathbf{R}}_{\mathbf{H}}^{ - 1} } \right)^{ - 1} } \right] \notag\\
 = &\mathbb{E}\left[ {\left( {\frac{{\tau _{\text{T}} P_{\text{T}} }}
{{N_0 }}{\mathbf{I}}_k  + {\mathbf{{\tilde E}_2\Lambda }}_{\mathbf{X}} {\mathbf{\tilde E}}_2^{\text{H}}  + {\mathbf{R}}_{\mathbf{H}}^{ - 1} } \right)^{ - 1} } \right] \notag\\
\approx &\left( {\frac{{\tau _{\text{T}} P_{\text{T}} }}
{{N_0 }}{\mathbf{I}}_k  + \frac{{\tau _{\text{D}} P_{\text{D}} }}
{{\Delta S_{\mathbf{X}}  + N_0 }}{\mathbf{\tilde E}}_{2}^{2} + {\mathbf{R}}_{\mathbf{H}}^{ - 1} } \right)^{ - 1}\notag\\
 = &{\text{diag}}\left( {\frac{{\beta _1^{\text{M}} }}
{{1 + \rho _1^{{\text{DA}}} \beta _1^{\text{M}} }}, \dots ,\frac{{\beta _K^{\text{M}} }}
{{1 + \rho _K^{{\text{DA}}} \beta _K^{\text{M}} }}} \right),
\end{align}
where
\begin{equation}\label{}
\rho _k^{{\text{DA}}}  = \frac{{\tau _{\text{T}} P_{\text{T}} }}
{{N_0 }} + \frac{{\tau _{\text{D}} P_{\text{D}} \left( {1 - 2{\text{BER}}_{vk} } \right)^2 }}
{{\Delta S_{\mathbf{X}}  + N_0 }},
\end{equation}
and the approximation above is due to the fact that the UL data length is usually long enough to hold that the two different data stream are uncorrelated. This means that ${\mathbf{\hat X\hat X}}^{\text{H}}  \to \tau _{\text{D}} P_{\text{D}} {\mathbf{I}}_k$ when $\tau _{\text{D}}$ is large. Hence, utilizing the above results, we have
\begin{equation}\label{}
\mathbb{E}\left[ {\left\| {{\mathbf{\hat h}}_k^{{\text{MMSE}}}  - {\mathbf{h}}_k } \right\|^2 } \right] = \frac{{M\beta _k^{\text{M}} }}
{{1 + \rho _k^{{\text{DA}}} \beta _k^{\text{M}} }}.
\end{equation}

Finally, the NMSE for the data-aided method with the MMSE estimator can be obtained based on the definition of NMSE
\begin{equation}\label{}
\mathcal{J}_k^{{\text{MMSE}}}
%= 10\log _{10} \left( {\frac{{\mathbb{E}\left[ {\left\| {{\mathbf{\hat h}}_k^{{\text{MMSE}}}  - {\mathbf{h}}_k } \right\|^2 } \right]}}
%{{\mathbb{E}\left[ {\left\| {{\mathbf{h}}_k } \right\|^2 } \right]}}} \right)
%&= 10\log _{10} \left( {\frac{{M\beta _k^{\text{M}}  - M\left( {\beta _k^{\text{M}} } \right)^2 \left( {\frac{1}
%{{\beta _k^{\text{M}} }} - \frac{1}
%{{\beta _k^{\text{M}} }}\frac{{\beta _k^{\text{M}} }}
%{{1 + \rho _k^{{\text{DA}}} \beta _k^{\text{M}} }}\frac{1}
%{{\beta _k^{\text{M}} }}} \right)}}
%{{M\beta _k^{\text{M}} }}} \right) \\
= 10\log _{10} \left( {\frac{1}
{{1 + \rho _k^{{\text{DA}}} \beta _k^{\text{M}} }}} \right),
\end{equation}
which proves \emph{Proposition 2}.
\end{IEEEproof}

\bibliographystyle{IEEEtran}
\bibliography{IEEEabrv,ref}

% that's all folks
\end{document}